\documentclass[fleqn,usenatbib]{mnras}

\usepackage{newtxtext,newtxmath}

\usepackage[T1]{fontenc}
\usepackage{ae,aecompl}

\usepackage{graphicx}   
\usepackage{amsmath}    
\usepackage{amssymb}    

\newcommand\grad{{\bmath\nabla}}
\newcommand\bcdot{{\bmath\cdot}}

\newcommand\vecxi{{\bmath\xi}}

\newcommand\vecu{{\bmath u}}

\newcommand\dd{\mathrm{d}}
\newcommand\DD{\mathrm{D}}

\newcommand\ii{\mathrm{i}}

\newcommand\f{\frac}
\newcommand\p{\partial}
\newcommand\cst{\mathrm{constant}}

\usepackage{xcolor}

\title[VSI as a travelling wave I]
{The vertical shear instability in protoplanetary discs as an outwardly travelling wave. I. Linear theory}

\author[G.\ I.\ Ogilvie, H.\ N.\ Latter and G.\ Lesur]
{Gordon I.\ Ogilvie,$^1$\thanks{E-mail: gio10@cam.ac.uk}, Henrik N.\ Latter$^{1}$ and Geoffroy Lesur$^{2}$
\\
$^{1}$Department of Applied Mathematics and Theoretical Physics,
University of Cambridge, Centre for Mathematical Sciences,\\
Wilberforce Road, Cambridge CB3 0WA, UK\\
$^{2}$Universit\'e Grenoble Alpes, CNRS, IPAG, 38000 Grenoble, France
}

\date{Accepted 2025 January 21. Received 2025 January 14; in original form 2024 November 11}

\pubyear{2025}

\begin{document}
\label{firstpage}
\pagerange{\pageref{firstpage}--\pageref{lastpage}}
\maketitle

\begin{abstract}
We revisit the global linear theory of the vertical shear instability (VSI) in protoplanetary discs with an imposed radial temperature gradient. We focus on the regime in which the VSI has the form of a travelling inertial wave that grows in amplitude as it propagates outwards. Building on previous work describing travelling waves in thin astrophysical discs, we develop a quantitative theory of the wave motion, its spatial structure and the physical mechanism by which the wave is amplified. We find that this viewpoint provides a useful description of the large-scale development of the VSI in global numerical simulations, which involves corrugation and breathing motions of the disc. We contrast this behaviour with that of perturbations of smaller scale, in which the VSI grows into a nonlinear regime in place without significant radial propagation.
\end{abstract}




\section{Introduction}

Over the last 15 years, scientific consensus has converged on a picture of protoplanetary discs in which the magnetorotational instability is mostly absent, because of insufficient ionisation, and instead accretion is driven by laminar non-ideal magnetic winds \citep[e.g.,][]{Turner14, Lesur2021}. Concurrently, researchers have better appreciated that protoplanetary discs are subject to a fascinating array of hydrodynamic instabilities, which may supply a low level of turbulent activity and/or form structures, such as zonal flows and vortices \citep{Lesur2023}. While probably unimportant for accretion, these instabilities are likely to influence dust diffusion and coagulation, and thus planet formation generally.

Researchers have concentrated  on the vertical shear instability \citep[VSI;][]{2013MNRAS.435.2610N}, especially, because of its relative robustness and supposed prevalence over several tens of au \citep{Pfeil2019, Lyra2019}. Current research activity is focused on adding more and more physical processes \citep[e.g.][]{2014A&A...572A..77S, 2016A&A...594A..57S, 2020ApJ...897..155F, 2020ApJ...891...30C, 2023A&A...670A.135Z}, and yet the VSI's fundamental dynamics are still incompletely understood. This uncertainty includes (unusually) its linear theory and initial growth mechanism, not only its nonlinear saturation.    

The VSI's local Boussinesq linear theory is satisfying and complete, both mathematically and physically \citep{1998MNRAS.294..399U, 2018MNRAS.474.3110L}, but it does not join up easily to the linear problem in vertically stratified local or global models \citep{2013MNRAS.435.2610N, 2015MNRAS.450...21B}. For example, the `body modes' of stratified models (growing inertial waves) fail to appear in the Boussinesq approximation at all, while the identification of the `surface modes' as Boussinesq modes remains insecure.
Moreover, we do not have a physical picture of how the VSI drives the growth of the `body modes'. 

The VSI's nonlinear behaviour throws up further puzzles. For example: Why are the (faster growing) surface modes suppressed and supplanted by the body modes? Why do extended trains of `corrugation waves' (body modes with vertical mode number $n=1$) ultimately dominate simulations and what selects their frequency? What saturates the growth of the corrugation waves -- is it controlled by their radial propagation through the disc (a linear process) or is it due to secondary instability \citep{2018MNRAS.474.3110L, 2022MNRAS.512.1639C}? Corrugation waves steepen as they travel; what limits the sharpness of their shear layers? Finally, VSI simulations break up into distinct wave zones \citep{2014A&A...572A..77S, 2022MNRAS.514.4581S}; what process selects which, of the several possible waves (of different frequency and $n$), dominates each zone?

This is the first of a series of papers that addresses some of these issues, employing analytical techniques complemented by carefully calibrated numerical experiments. Our main goal is to develop a linear, and weakly nonlinear, theory for travelling VSI body modes in global disc models.
We find that larger-scale modes, with vertical wavenumber $n=1,2$, travel radially outwards as they grow; they therefore propagate away from their birthplace to radii with different disc properties, which then impact on any further growth and continuing propagation. This behaviour contrasts with that of smaller-scale modes (of higher $n$), which grow and saturate in place without significant radial propagation.
 As nonlinear VSI simulations are dominated by outwardly travelling perturbations, it is essential to understand them. This paper outlines the linear theory of VSI travelling waves, superseding previous local analyses, which were unable to track their global propagation, and previous global analyses, which were limited to standing waves and relatively short radial extents. Ensuing papers will explore the VSI's weakly nonlinear interactions, which govern the transition between wave zones, and present illustrative numerical simulations. 

There are several new results in this paper. We provide a novel physical explanation for the VSI when it takes the form of a travelling inertial wave; the growth mechanism can be understood either in terms of the work done on the elliptical fluid circuits that constitute the basic wave motion, or in terms of Reynolds stresses working on both the vertical and radial shears. We find that the $r\phi$ Reynolds stress is surprisingly important and accounts for the majority of the energy budget of the VSI. We also demonstrate that steady linear wavetrains, involving `corrugation' and `breathing' modes, are an inevitable outcome of the VSI, if there is a continuous supply of small-amplitude fluctuations at small radii. However, VSI modes with larger vertical mode numbers, or modes seeded at smaller radii, grow in place and must saturate through a nonlinear mechanism.

The paper is structured as follows. In Section~\ref{s:background}, we provide the necessary background from which we can start constructing the theory, most importantly material from `discoseismology'. In fact, understanding the internal fluid motions of waves in discs enables us to devise a relatively transparent first explanation for the growth of the body modes (Section~\ref{s:meridional}). We outline our (standard) global disc model and underlying assumptions in Section~\ref{s:model}, as well as discussing the linearized equations and the associated energy equations that govern the growth of perturbations. Then, in Section~\ref{s:travelling}, we present an analytical description of travelling waves and their growth due to the VSI. In Section~\ref{s:connecting} we contrast this behaviour with the local growth without propagation, found in previous work, and we discuss the connection between the two regimes. Our findings are discussed and summarized in Sections \ref{s:implications} and~\ref{s:conclusion}. Some of our more technical derivations are contained in the appendices.

\section{Background material}
\label{s:background}

The main purpose of this paper is to understand the VSI as a wave that grows as it travels outwards through a disc with an imposed radial temperature gradient; to do so, we must generalize the existing theory of travelling waves in adiabatic or isothermal discs. So as to set the scene, we first briefly summarize previous analyses of the VSI (Section~\ref{s:previous}), before reviewing the theory of wave propagation in discs (Section~\ref{s:waves}).

\subsection{Previous approaches to the VSI}
\label{s:previous}

\citet{1998MNRAS.294..399U} presented a linear analysis of axisymmetric perturbations to a disc model that was baroclinic, involving a vertical shear, as suggested by previous calculations of the detailed structure of viscous discs. They adopted the Boussinesq approximation, which excludes sound waves and is appropriate for disturbances that are slow and of short wavelength compared to the scale-height of the disc. Their analysis generally included a weak magnetic field, but in the non-magnetic case they discovered the VSI as an application to accretion discs of the instability that \citet{1967ApJ...150..571G} and \citet{1968ZA.....68..317F} had discussed in radiative zones of differentially rotating stars, where the stabilizing effect of buoyancy is negated by thermal diffusion acting on perturbations of sufficiently short wavelength. By obtaining a local dispersion relation for plane-wave disturbances, \citet{1998MNRAS.294..399U} deduced that instability occurs when the wavevector is slightly inclined from the radial direction, giving a maximum growth rate comparable to the rate of vertical shear. They suggested that the VSI could give rise to anisotropic turbulence \citep[see also][]{2003A&A...404..397U}.

\citet{2013MNRAS.435.2610N} rediscovered the VSI in the context of numerical simulations of accretion discs with an imposed radial temperature gradient, motivated by models of protostellar or protoplanetary discs that are heated by stellar radiation. In addition to carrying out the first global axisymmetric simulations of the VSI, they presented a new linear analysis of the instability. They derived a reduced model that is local in the radial direction but includes the full vertical structure of the disc; it excludes sound waves but is anelastic (retaining the non-uniform density structure) rather than Boussinesq, and is also geostrophic (balancing the radial Coriolis force with a pressure gradient), appropriate for low-frequency inertial waves associated with the rotation of the disc. This model allowed them to compute normal modes with a wavelike structure in the radial direction and having a growth rate and a non-trivial vertical structure determined by an eigenvalue problem.

Numerical solutions of this eigenvalue problem produced `surface' and `body' modes that could be related to features seen in the nonlinear numerical simulations. The surface modes have a higher growth rate but are concentrated near the vertical boundaries and their properties depend on the boundary conditions; they can be related to the very small-scale disturbances that grow at an early stage in the simulations. The body modes, which include `corrugation' and `breathing' modes with opposite symmetry about the midplane, are more robust predictions of the model and include the `fundamental corrugation mode' (labelled as $n=1$ in the present paper) that is seen to dominate at late times in the simulations. However, the authors were unable to explain why higher-order body modes with larger growth rates were not preferred. \citet{2013MNRAS.435.2610N} did mention that the fundamental corrugation mode could be either an inwardly propagating inertial wave or an outwardly propagating acoustic wave; note that this refers to the direction of phase propagation rather than that of the group velocity, which is more physically meaningful and is outward in the case of the relevant inertial waves.

\citet{2015MNRAS.450...21B} found analytical solutions (involving Hermite polynomials of complex argument) for the modes of the reduced model derived by \citet{2013MNRAS.435.2610N} in a vertically isothermal disc without vertical boundaries. The complex, analytical dispersion relation (their equation 34) gives both the frequency and the growth rate of the mode in terms of the radial wavenumber, vertical mode number (similar to $n$ in the present paper) and the radial temperature gradient. This dispersion relation explains the local growth of the VSI body modes and shows that they are also travelling inertial waves that grow because of the vertical shear. The model predicts that the growth rate is proportional to $\sqrt{n}$, but is expected to break down when $n$ is too large. \citet{2015MNRAS.450...21B} also found surface modes when the model was truncated with artificial vertical boundaries, or when a polytropic disc with a definite surface was used instead. They further computed the 2D structure and growth rate of global axisymmetric modes in a locally isothermal, compressible disc model, finding body modes in the form of standing inertial waves (in both radial and vertical directions) with Lindblad resonances close to or beyond the outer radius of the domain.

\citet{2015ApJ...811...17L} also computed the growth rates and vertical structures of surface and body modes with a wavelike radial structure in a radially local disc model, focusing mainly on the effects of more realistic thermal physics, but they did describe the VSI as an unstable inertial wave.

The idea that the VSI gives rise to a train of outwardly propagating inertial waves was stated most clearly by \citet{2022MNRAS.514.4581S}, based on careful analysis of global axisymmetric simulations in a domain of large radial extent. In retrospect, this property is evident in the earlier work of \citet{2013MNRAS.435.2610N} and \citet{2014A&A...572A..77S}. Several other VSI simulations have had too small a radial domain to allow the wavetrain to emerge. \citet{2022MNRAS.514.4581S} also noted the partitioning of the disc into discrete wave zones, each spanning a factor of approximately two in radius (or three in orbital frequency). Within each wave zone the travelling wave has a coherent frequency and its radial wavelength follows the appropriate dispersion relation. A slight overlap between adjacent wave zones appears to allow a handover from the inner wave to its lower-frequency successor.
Most recently, \citet{2024MNRAS.529..918D} have carried out a resolution study and found that a higher spatial resolution favours lower-frequency inertial waves of shorter wavelength, which can have a more irregular evolution as they propagate outwards, although the duration of these simulations is shorter than those of \citet{2022MNRAS.514.4581S}.

\subsection{Axisymmetric waves in an isothermal disc}
\label{s:waves}

To help us make sense of the wavelike properties of the VSI, we review some earlier work on wave propagation in astrophysical discs.

\subsubsection{The disc as a waveguide}

The radial propagation of axisymmetric waves in an accretion disc was studied by \citet{1993ApJ...409..360L}, who considered a thin, vertically isothermal disc orbiting in a central potential, in the absence of self-gravity, magnetic fields and viscosity. [Aspects of the same problem for a polytropic disc had earlier been treated by \citet{1986AcA....36...43L}.] They showed that the disc acts a waveguide, allowing certain wave modes with distinct vertical structures to propagate radially through the disc.

In such a wave, the perturbation $X'$ of a fluid variable $X$, which could be density, pressure or one of the velocity components in cylindrical polar coordinates $(r,\phi,z)$, has the form
\begin{equation}
  X'=\text{Re}\left\{\tilde X'(r,z)\,\exp\left[-\ii\omega t+\ii\int k(r)\,\dd r\right]\right\}.
\label{wkb}
\end{equation}
The phase factor given by the exponential part of this expression describes a wave that travels in the radial direction. The wave has an angular frequency $\omega$, independent of $r$, and a radial wavenumber $k(r)$, dependent on $r$. The prefactor $\tilde X'(r,z)$ is a wave amplitude that varies typically on scales comparable to the dimensions of the disc.

Each wave mode corresponds to a different branch of the dispersion relation $\omega(k)$, relating the angular frequency of the wave to its radial wavenumber. The dispersion relation also depends on $r$ because the properties of the disc vary in the radial direction; in order to satisfy the dispersion relation at all radii within some range, the wavenumber $k$ must depend on $r$.

This description is valid when $kr\gg1$, i.e.\ when the radial lengthscale $1/k$ associated with the wave is short compared to the radial distance $r$ from the centre, which is the characteristic scale on which the properties of a smooth disc vary. Typically $1/k$ may be comparable to, or shorter than, the vertical scaleheight $H\ll r$ of the disc.

The solutions have the form of standing waves in the vertical direction as they are naturally contained by the vertically localized density distribution, even though a vertically isothermal disc lacks a definite surface at which the density goes to zero. In the radial direction, the solutions have the form of travelling waves with phase velocity $\omega/k$ and group velocity $\dd\omega/\dd k$. The high-frequency branches have the character of acoustic waves, in which the predominant restoring force is the gradient of pressure perturbations related to the compression of the fluid, while the low-frequency branches have the character of inertial waves, in which the predominant restoring force is the Coriolis force, modified by orbital shear.

Since the real part of the complex conjugate of a complex number is identical to the real part of the same number, there is no loss of generality in assuming that $\omega\ge0$, and we adopt this assumption throughout this paper.

\subsubsection{Wave modes in a strictly isothermal disc}

The simplest physical situation occurs when the dynamical response of the gas is isothermal, as well as its equilibrium vertical structure. Then buoyancy forces are absent and the dispersion relation for a Keplerian disc takes the form
\begin{equation}
  (\omega^2-n\Omega^2)(\omega^2-\Omega^2)=(\omega c_\text{s}k)^2,
\label{dispersion}
\end{equation}
where $\Omega$ is the Keplerian angular velocity, $c_\text{s}$ is the sound speed (related to the vertical scaleheight $H$ by $c_\text{s}=H\Omega$) and $n=0,1,2,\dots$ is the vertical mode number. [The mode number $n$ used by \citet{1993ApJ...409..360L} corresponds to our $n-1$. The dispersion relation~(\ref{dispersion}) was already given by \citet{1987PASJ...39..457O}, whose focus was on relativistic discs.]

The vertical structure of the modes is generally oscillatory, with the horizontal velocity perturbations (or displacements) being proportional to a polynomial of degree~$n$ in~$z$, with $n$ zeros, while the vertical velocity perturbation (or displacement) is proportional to a polynomial of degree $n-1$. For each value of $n$ there is a pair of wave modes, corresponding to the two roots of the quadratic equation~(\ref{dispersion}) for $\omega^2$.

\begin{figure*}
    \centering
    \includegraphics[width=18cm]{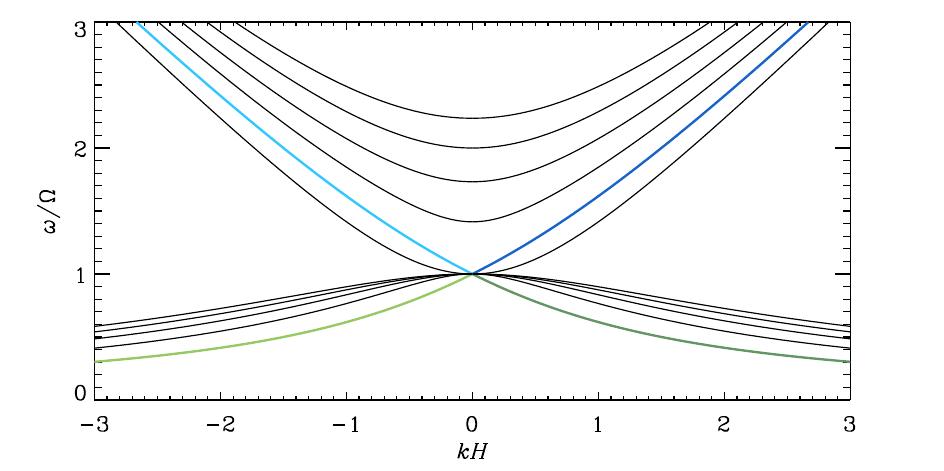}
    \caption{Part of the dispersion relation~(\ref{dispersion}) for a vertically isothermal Keplerian disc. The high-frequency branches shown here with $\omega>\Omega$ have $n=0,1,\dots5$, where $\omega$ increases with $n$. The low-frequency branches shown here with $\omega<\Omega$ have $n=1,2,\dots,5$, where again $\omega$ increases with $n$. The coloured branches have $n=1$ and are referred to in Section~\ref{s:neq1}.}
    \label{f:waves}
\end{figure*}

Part of the dispersion relation is illustrated in Fig.~\ref{f:waves}. A notable feature is the point $(kH,\omega/\Omega)=(0,1)$ through which many of the branches pass. 
This corresponds to both the Lindblad resonance and the vertical resonance of the disc. (Keplerian orbits are special in that small perturbations either within or out of the plane of the original orbit lead to oscillations at the orbital frequency, resulting in closed perturbed orbits.)

\subsubsection{Modes with $n=0$}

Two pairs of modes are of particular interest. The case $n=0$ involves a purely horizontal motion independent of $z$, and has either
\begin{equation}
  \omega^2=\Omega^2+c_\text{s}^2k^2\qquad\text{or}\qquad\omega^2=0.
\end{equation}
The first possibility corresponds to the density wave, which is an inertial--acoustic wave, as also occurs in a 2D disc model with no vertical structure. The second possibility corresponds to a stationary vortical mode, in which the azimuthal velocity is perturbed in the form of a zonal flow with an oscillatory radial structure, while pressure perturbations ensure a geostrophic force balance.

Although the density wave is not involved in the VSI, one of the results of this paper (see Section~\ref{s:flux_steady}) is a relationship between the non-conservative propagation of the density wave in a locally isothermal disc, as highlighted in recent work on planet--disc interactions \citep{2016ApJ...832..166L,2020ApJ...892...65M}, and the growth of travelling inertial waves due to the VSI.

\subsubsection{Modes with $n=1$}
\label{s:neq1}

The case $n=1$, which has been found to play a dominant role in the VSI, involves a vertical motion independent of $z$ together with a horizontal motion proportional to $z$. It has
\begin{equation}
  \f{\omega}{\Omega}=\f{1}{2}\left(\pm|kH|+\sqrt{4+(kH)^2}\right).
\end{equation}
This pair of branches forms a cross at the point $(kH,\omega/\Omega)=(0,1)$ in Fig.~\ref{f:waves} (coloured curves), where their gradients are $\pm1/2$.

The high-frequency branch given by the $+$ sign is a predominantly acoustic wave. As $k$ increases from $0$ to $+\infty$ (dark blue curve), $\omega$ increases from $\Omega$ to $+\infty$. The radial group velocity
\begin{equation}
  \f{\dd\omega}{\dd k}=\f{1}{2}\left(1+\f{kH}{\sqrt{4+(kH)^2}}\right)c_\text{s}
\end{equation}
is positive, indicating an outwardly propagating wave, and increases from $c_\text{s}/2$ to $c_\text{s}$ as the wave becomes more acoustic in character. The corresponding high-frequency branch with $k<0$ (light blue curve) has a negative group velocity and represents an inwardly propagating wave.
 
The low-frequency branch given by the $-$ sign is a predominantly inertial wave and is of special interest in this paper. As $k$ increases from $0$ to $+\infty$ (dark green curve), $\omega$ decreases from $\Omega$ to $0$. The radial group velocity
\begin{equation}
  \f{\dd\omega}{\dd k}=-\f{1}{2}\left(1-\f{kH}{\sqrt{4+(kH)^2}}\right)c_\text{s}
\end{equation}
is negative, indicating an inwardly propagating wave, and decreases from $c_\text{s}/2$ to $0$. The corresponding low-frequency branch with $k<0$ (light green curve) has a positive group velocity and represents an outwardly propagating wave. Note that the phase velocity $\omega/k$ of the inertial wave has the opposite sign to the group velocity.

\subsubsection{Radial propagation of inertial waves}

In general, inertial waves are able to propagate radially either inwards or outwards, provided that $\omega<\Omega$. Since $\omega$ is constant and $\Omega$ decreases outwards, inertial waves are able to propagate interior to the Lindblad resonance $r=R$ at which $\omega=\Omega$. Modes with $n>1$ have a turning point at the Lindblad resonance and are reflected there. The $n=1$ mode in a Keplerian disc is special and is able to propagate smoothly through the Lindblad resonance, appearing as an acoustic wave in the region $r>R$ where $\omega>\Omega$.

The $n=1$ mode is also unique in having a vertical displacement that is independent of $z$, which means that it causes a corrugation or bending of the midplane of the disc. It can therefore be thought of as a corrugation or bending wave, as well as an inertial (or acoustic) wave.

Previous analyses of the radial variation of the amplitudes of travelling waves generally rely on the conservation of wave action, which in turn is related to the conservation of energy. For example, \citet{1998ApJ...504..983L} relate the radial flux of wave action to the square of the wave amplitude for adiabatic perturbations in an inviscid disc. In the case of a steady, unforced wavetrain, constancy of the flux determines the radial variation of the amplitude. The VSI requires a different approach because energy and wave action are not conserved in the presence of an imposed temperature gradient.

\subsubsection{Meridional circuits}
\label{s:meridional}

\begin{figure}
  \centering
  \includegraphics[width=8cm]{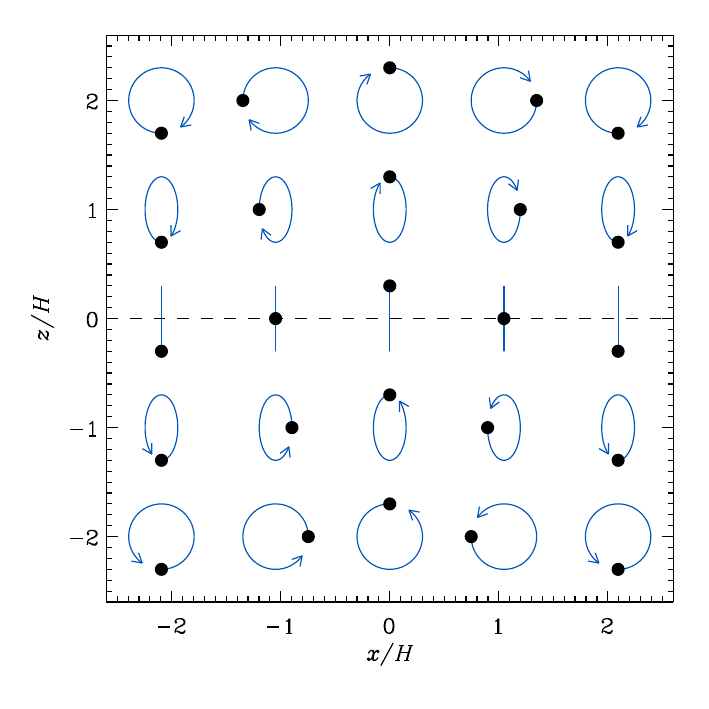}
  \caption{Motion of fluid elements in the meridional plane in a small radial section of an outwardly propagating inertial wave. The case shown here has $n=1$ and $\omega=\Omega/2$, such that the radial wavelength is $2\pi/k=(4\pi/3)H$, and an amplitude $\mathcal{A}=0.3$ is chosen for illustrative purposes. The variable $x$ is a local radial coordinate. The black dots denote the displaced locations of selected fluid elements at a particular instant of time. Over one wave period, each fluid element follows the elliptical path indicated by the blue arrow. The motion is clockwise above the midplane, anticlockwise below the midplane and vertical at the midplane (indicated by a dashed line).}
  \label{f:ellipses}
\end{figure}

The motion in the meridional $(r,z)$ plane can be visualized by considering the Lagrangian displacement $\vecxi$, which is the (small) change in the position of a fluid element due to the presence of the wave. In the case of the $n=1$ modes, the relation
\begin{equation}
  \f{\xi_r}{\xi_z}=\ii kz\left(\f{\omega^2}{\omega^2-\Omega^2}\right)
\end{equation}
 can be obtained by comparing the radial and vertical components of the linearized equation of motion and eliminating the pressure perturbation (see Section~\ref{s:structure} and Appendix~\ref{s:asymptotic}). If we consider an outwardly propagating inertial wave ($k<0$ and $0<\omega<\Omega$) then we see that $\xi_r/\xi_z=\ii az$ for some $a>0$. Above the midplane ($z>0$), $\xi_r$ therefore lags $\xi_z$ by one quarter of a wave period, since our sign convention is that the phase of the wave decreases with time. This means that the fluid elements move in elliptical paths in a clockwise direction in the $(r,z)$ plane. Below the midplane the motion is an anticlockwise ellipse and at the midplane the motion is purely vertical. This behaviour is illustrated in Fig.~\ref{f:ellipses}.

In travelling wave modes with $n>1$, the motion in the meridional plane is also elliptical, but changes direction between clockwise and anticlockwise at certain heights. Nevertheless, in a sense that we will quantify later, the motion above the midplane in an outwardly propagating inertial wave is more clockwise than anticlockwise.

The fact that the motion is elliptical relies on the assumption that the wave is standing in the vertical direction and travelling in the radial direction. In other words, there has been time for the wave to sense the vertical extent of the disc and to be contained and guided by its vertical stucture without yet having sensed the radial boundaries of the disc. In contrast, if the VSI is studied in the Boussinesq approximation (without vertical structure and with periodic vertical boundary conditions), such solutions do not naturally emerge. Instead, waves travel both vertically and radially, and fluid elements oscillate back and forth along a tilted \emph{line} in the meridional plane, not around an elliptical circuit.

\subsubsection{The instability mechanism of VSI body modes}

The preceding subsection provides the conceptual tools to understand the growth of inertial waves (body modes) due to the VSI. While the instability mechanism of the monotonically growing Boussinesq VSI modes (or surface modes) is similar to classical centrifugal instability \citep[see, e.g.,][]{2018MNRAS.474.3110L}, the mechanism for the wavelike body modes is somewhat different.  

\begin{figure}
  \centering
  \includegraphics[width=2.8cm]{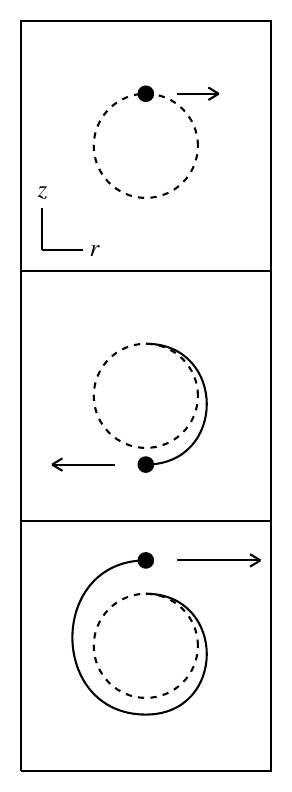}
  \caption{Illustration of the motion of a typical fluid element above the midplane in an outwardly propagating inertial wave. The dashed line represents the clockwise elliptical circuit of the fluid element in the absence of vertical shear (cf.\ Fig.~\ref{f:ellipses}). The three panels represent a time-sequence. The arrows indicate additional centrifugal accelerations due to the vertical angular momentum gradient (equation~\ref{f_r}), which do positive work on the wave. In the top panel, the fluid element has a slightly greater specific angular momentum than the background and experiences an excess centrifugal force. The opposite is true in the middle panel.}
  \label{f:cartoon}
\end{figure}

In a protoplanetary disc with an imposed temperature profile $T(r)$ such that $\dd T/\dd r<0$, the equilibrium state of the disc is baroclinic and the angular velocity $\Omega$ depends slightly on $z$, having a peak at the midplane (see Section~\ref{s:equilibrium}). This weak vertical shear slightly perturbs the picture of wave propagation presented so far. Consider a fluid element executing a clockwise elliptical path above the midplane in an outwardly propagating inertial wave (Fig.~\ref{f:cartoon}). The fluid element still preserves its specific angular momentum $l$, but because the equilibrium value $l=r^2\Omega$ slightly decreases with $z$ above the midplane, the fluid element has an excess of angular momentum relative to the background in the upper half of the circuit and a deficit in the lower half. It therefore experiences an excess centrifugal force in the upper half and a deficit in the lower half. These perturbing forces (the arrows in Fig.~\ref{f:cartoon}) are in tune with the radial motion in the wave and therefore do positive work on it, so amplifying the motion. If the wave is confined in a shearing box with periodic radial boundary conditions, then its amplitude will grow in time, leading to the spiralling paths shown in Fig.~\ref{f:cartoon}. On the other hand, if the wave is not confined, then the added energy will be carried radially outwards by the travelling wave.

Lastly, it is now clearer why growing body modes fail to appear in Boussinesq models. Simply, Boussinesq inertial waves comprise radial velocity oscillations that are out of phase with the excess centrifugal force: the excess force is greatest at the oscillation's maximum meridional displacement, but at this point the radial velocity is zero; on the other hand, at the point of maximum radial velocity, the displacement is zero and so is the excess force. When instability does occur in Boussinesq models, it is monotonic: a fluid blob's poloidal displacement is simply reinforced and keeps growing without oscillation.

\section{Disc model and governing equations}
\label{s:model}

In this section we define the physical and mathematical model with which we work throughout this paper. The model is similar or identical to most previous treatments of the VSI and is intentionally simplified to highlight the essential physics. We include some novel analysis to help with the physical interpretation.

\subsection{Basic equations}
\label{s:basic}

The central star is modelled as a point mass $M$ with gravitational potential $\Phi=-GM/\sqrt{r^2+z^2}$, where again $(r,\phi,z)$ are cylindrical polar coordinates. We assume that the disc is composed of a non-self-gravitating, ideal fluid that is `locally isothermal', such that its pressure $p$ is related to its density $\rho$ by $p=c_\text{s}^2\rho$, with a sound speed $c_\text{s}(r)$ that depends on the cylindrical radius.

The locally isothermal model, which has been adopted in innumerable studies of protoplanetary discs, has an imposed radial temperature profile $T(r)$, with $T\propto c_\text{s}^2$, that is dictated by an external radiation field due to the central star and naturally decreases outwards. It corresponds to an idealized situation in which the thermal physics of the disc allows it to adjust instantaneously to this temperature profile, counteracting any adiabatic heating or cooling resulting from compressive motions of the gas.

Throughout this paper, we assume the dynamics to be axisymmetric. The remaining governing equations for the disc are the three components of the equation of motion,
\begin{align}
&  \f{\DD u_r}{\DD t}-\f{u_\phi^2}{r}=-\f{\p\Phi}{\p r}-\f{1}{\rho}\f{\p p}{\p r}, \\
&  \f{1}{r}\f{\DD(ru_\phi)}{\DD t}=0, \\
&  \f{\DD u_z}{\DD t}=-\f{\p\Phi}{\p z}-\f{1}{\rho}\f{\p p}{\p z},
\end{align}
and the equation of mass conservation,
\begin{equation}
  \f{\p\rho}{\p t}+\f{1}{r}\f{\p(r\rho u_r)}{\p r}+\f{\p(\rho u_z)}{\p z}=0,   
\end{equation}
where
\begin{equation}
  \f{\DD}{\DD t}=\f{\p}{\p t}+u_r\f{\p}{\p r}+u_z\f{\p}{\p z}
\end{equation}
is the Lagrangian time-derivative following the fluid motion and $(u_r,u_\phi,u_z)$ are the velocity components in cylindrical coordinates.

\subsection{Energy equation}

It can be helpful for physical interpretation to construct the energy equation implied by the basic equations above. In fact, to understand how the VSI operates, we are particularly interested in the extent to which the energy of the disc, or of a wave within it, is \textit{not} conserved.

An equation for the mechanical (kinetic plus gravitational) energy is
\begin{align}
  &\f{\p}{\p t}\left[\rho\left(\f{1}{2}u^2+\Phi\right)\right]+\f{1}{r}\f{\p}{\p r}\left[r\rho u_r\left(\f{1}{2}u^2+\Phi\right)+rpu_r\right]\nonumber\\
  &\qquad+\f{\p}{\p z}\left[\rho u_z\left(\f{1}{2}u^2+\Phi\right)+pu_z\right]=p\left[\f{1}{r}\f{\p(ru_r)}{\p r}+\f{\p u_z}{\p z}\right].
\label{energy_mechanical}
\end{align}
This equation is in conservative form, involving the time-derivative of an energy density and the divergence of an energy flux density, but it also includes a source (or sink) term on the right-hand side in the form of the $P\,\dd V$ work of pressure forces.

Under the locally isothermal conditions considered in this paper, the relevant thermal energy per unit mass is $c_\text{s}^2(\ln\rho+\cst)$. (See Appendix~\ref{s:energy} for an explanation and interpretation of this result.) The total energy equation is
\begin{align}
  &\f{\p}{\p t}\left[\rho\left(\f{1}{2}u^2+\Phi+c_\text{s}^2\ln\rho\right)\right]\nonumber\\
  &\qquad+\f{1}{r}\f{\p}{\p r}\left[r\rho u_r\left(\f{1}{2}u^2+\Phi+c_\text{s}^2\ln\rho\right)+rpu_r\right]\nonumber\\
  &\qquad+\f{\p}{\p z}\left[\rho u_z\left(\f{1}{2}u^2+\Phi+c_\text{s}^2\ln\rho\right)+pu_z\right]=(\rho\ln\rho)u_r\f{\dd c_\text{s}^2}{\dd r}.
\label{energy_non-conservation}
\end{align}
Once again, conservation is broken, with the source term on the right-hand side of this equation involving radial motions that sample the imposed temperature gradient in the disc. This source term can be interpreted physically as the effective cooling or heating that is needed to maintain the locally isothermal condition.

\subsection{Equilibrium balances}
\label{s:equilibrium}

We suppose that the disc falls into a steady equilibrium state characterized by purely orbital motion, so that $u_r=u_z=0$, while $u_\phi=r\Omega(r,z)$. In addition, we prescribe the radial profiles of midplane density and sound speed. The vertical density structure of the disc and its angular velocity follow from the radial and vertical force balances
\begin{align}
  -r\Omega^2&=-\f{\p\Phi}{\p r}-\f{1}{\rho}\f{\p p}{\p r}, \label{eqm1}\\
  0&=-\f{\p\Phi}{\p z}-\f{1}{\rho}\f{\p p}{\p z}. \label{eqm2}
\end{align}
Before giving the solution of these equations, we note that eliminating $\Phi$ between them by cross-differentiation (equivalent to forming the vorticity equation by taking the curl of the equation of motion) leads to the thermal-wind equation
\begin{align}
  -r\f{\p\Omega^2}{\p z}&=\f{1}{\rho^2}\left(\f{\p p}{\p r}\f{\p\rho}{\p z}-\f{\p p}{\p z}\f{\p\rho}{\p r}\right)\nonumber\\
  &=\f{\dd c_\text{s}^2}{\dd r}\f{\p\ln\rho}{\p z}\label{twe}\\
  &=-\f{\dd\ln T}{\dd r}\f{\p\Phi}{\p z}.\nonumber
\end{align}
This relates the vertical shear to the baroclinicity of the disc and to the imposed radial temperature gradient. A baroclinic state is one in which the gradients of the thermodynamic variables are not aligned. In the present situation, the temperature gradient, which is purely in the radial direction, is not aligned with the density gradient, which has a vertical component.

An explicit solution of the equilibrium equations is given by \citep[cf.][]{2013MNRAS.435.2610N}
\begin{align}
  &\rho=\rho_\text{m}\exp\left(\f{\Phi_\text{m}-\Phi}{c_\text{s}^2}\right),\\
  &r\Omega^2=\f{\dd\Phi_\text{m}}{\dd r}+c_\text{s}^2\f{\dd\ln\rho_\text{m}}{\dd r}+\left(\Phi-\Phi_\text{m}+c_\text{s}^2\right)\f{\dd\ln T}{\dd r},
\end{align}
in which $\Phi_\text{m}(r)=-GM/r$ is the potential in the midplane and $\rho_\text{m}(r)$ is the density in the midplane. In the expression for $r\Omega^2$ given here, the terms involving $c_\text{s}^2$ cause small departures of the angular velocity from the Keplerian value
\begin{equation}
  \Omega_\text{K}(r)=\left(\f{GM}{r^3}\right)^{1/2}.
\end{equation}
The midplane density $\rho_\text{m}$ (or, equivalently, the surface density of the disc) can be chosen freely.

For a thin disc we can approximate $\Phi-\Phi_\text{m}$ as $\f{1}{2}\Omega_\text{K}^2z^2$. Then the vertical profile of the density is Gaussian,
\begin{equation}
  \rho=\rho_\text{m}\,\exp\left(-\f{z^2}{2H^2}\right),
\label{rho_thin}
\end{equation}
with scaleheight $H(r)$ given as usual by $H=c_\text{s}/\Omega_\text{K}$, while
\begin{equation}
  r\Omega^2=r\Omega_\text{K}^2+c_\text{s}^2\f{\dd\ln\rho_\text{m}}{\dd r}+\left(1+\f{z^2}{2H^2}\right)\f{\dd c_\text{s}^2}{\dd r}.
\label{omega_thin}
\end{equation}
The last term, proportional to $z^2$, gives the vertical shear at a rate compatible with the thermal-wind equation~(\ref{twe}).

The surface density $\Sigma(r)$ of the thin disc is related to the midplane density and scaleheight by $\Sigma=\int\rho\,\dd z=\sqrt{2\pi}\rho_\text{m}H$.

\subsection{Linearized equations}

Next, the basic state of Section~\ref{s:equilibrium} is perturbed by a small axisymmetric disturbance. In what follows, the basic state will be denoted by $\Omega$, $\rho$ and $p$, while the (Eulerian) perturbations of physical quantities are distinguished by primes. 

The linearized equations of Section~\ref{s:basic} are
\begin{align}
  &\f{\p u_r'}{\p t}-2\Omega u_\phi'=-\f{1}{\rho}\f{\p p'}{\p r}+\f{\rho'}{\rho^2}\f{\p p}{\p r}, \label{Eulin1}\\
  &\f{\p u_\phi'}{\p t}+u_r'\f{1}{r}\f{\p l}{\p r}+u_z'\f{1}{r}\f{\p l}{\p z}=0, \label{Eulin2}\\
  &\f{\p u_z'}{\p t}=-\f{1}{\rho}\f{\p p'}{\p z}+\f{\rho'}{\rho^2}\f{\p p}{\p z}, \label{Eulin3}\\
  &\f{\p\rho'}{\p t}+\f{1}{r}\f{\p(r\rho u_r')}{\p r}+\f{\p(\rho u_z')}{\p z}=0, \label{Eulin4}
\end{align}
where $l=r^2\Omega$ is the specific angular momentum.

Instead of the velocity perturbations, however, it can be useful to work with the meridional displacements $(\xi_r,\xi_z)$, defined such that $u_r'=\p\xi_r/\p t$ and $u_z'=\p\xi_z/\p t$. We may then integrate equations (\ref{Eulin2}) and (\ref{Eulin4}) (assuming non-zero frequency) to find $u_\phi'$ and $\rho'$ in terms of $\xi_r$ and $\xi_z$. Thus
\begin{align}
  &u_\phi'=-\f{\xi_r}{r}\f{\p l}{\p r}-\f{\xi_z}{r}\f{\p l}{\p z},\label{uphip}\\
  &\rho'=-\f{1}{r}\f{\p(r\rho\xi_r)}{\p r}-\f{\p(\rho\xi_z)}{\p z},\label{rhop}\\
  &p'=c_\text{s}^2\rho'.\label{pp}
\end{align}
Equations (\ref{Eulin1}) and (\ref{Eulin3}) then become
\begin{align}
 &\f{\p^2\xi_r}{\p t^2}+\f{\xi_r}{r^3}\f{\p l^2}{\p r}+\f{\xi_z}{r^3}\f{\p l^2}{\p z}=-\f{\p}{\p r}\left(\f{p'}{\rho}\right)+\f{\dd c_\text{s}^2}{\dd r}\f{\rho'}{\rho},\label{d2xir}\\
 &\f{\p^2\xi_z}{\p t^2}=-\f{\p}{\p z}\left(\f{p'}{\rho}\right).\label{d2xiz}
\end{align}

\subsection{Non-conservation of wave energy}

To assist with the physical interpretation of the VSI, we examine to what extent the energy of the perturbations is not conserved. We derive two different, but related, forms of the wave energy equation; note that, in the absence of an exact conservation law, there is not a unique choice of a nearly conserved quantity.

Common to both approaches is that we multiply equation~(\ref{Eulin1}) by $\rho u_r'$, equation~(\ref{Eulin3}) by $\rho u_z'$ and equation~(\ref{Eulin4}) by $p'/\rho$, add the results and carry out some rearrangements, finding
\begin{align}
  &\f{\p}{\p t}\left[\f{1}{2}\rho\left(u_r'^2+u_z'^2\right)+\f{p'^2}{2p}\right]+\f{1}{r}\f{\p(ru_r'p')}{\p r}+\f{\p(u_z'p')}{\p z}\nonumber\\
  &\qquad=\f{\dd c_\text{s}^2}{\dd r}u_r'\rho'+2\Omega\rho u_r'u_\phi'.
\label{energy_wave}
\end{align}

In the first approach, we substitute expression~(\ref{uphip}) for $u_\phi'$ and incorporate the first term that it produces into the time-derivative to make
\begin{align}
 &\f{\p}{\p t}\left[\f{1}{2}\rho\left(u_r'^2+u_z'^2\right)+\f{\rho\xi_r^2}{2r^3}\f{\p l^2}{\p r}+\f{p'^2}{2p}\right]+\f{1}{r}\f{\p(ru_r'p')}{\p r}+\f{\p(u_z'p')}{\p z}\nonumber\\
 &\qquad=\f{\dd c_\text{s}^2}{\dd r}u_r'\rho'-\rho r\f{\p\Omega^2}{\p z}u_r'\xi_z.
\label{energy}
\end{align}
In this version, the wave energy density (in square brackets) includes the meridional part of the perturbation kinetic energy, the epicyclic potential energy associated with the radial gradient of angular momentum, and an acoustic contribution involving the pressure perturbation.
The right-hand side consists of two source terms that violate the conservation of wave energy, and thus may allow the wave to grow or decay. Both terms involve the radial velocity perturbation; the first couples the radial temperature gradient and the density perturbation, while the second couples the vertical shear and the vertical displacement. Using the thermal-wind equation~(\ref{twe}), the two source terms can be combined into the expression
\begin{equation}
  \left(\rho'+\xi_z\f{\p\rho}{\p z}\right)u_r'\f{\dd c_\text{s}^2}{\dd r},
\end{equation}
which is more closely (and yet not straightforwardly) related to the source term in the fully nonlinear energy equation~(\ref{energy_non-conservation}).

An equivalent way to think about the energy source is to rearrange equation~(\ref{d2xir}) in the form
\begin{equation}
  \f{\p^2\xi_r}{\p t^2}+\f{\xi_r}{r^3}\f{\p l^2}{\p r}=-\f{\p}{\p r}\left(\f{p'}{\rho}\right)+f_r,
\end{equation}
where
\begin{equation}
  f_r=\f{\dd c_\text{s}^2}{\dd r}\f{\rho'}{\rho}-\xi_zr\f{\p\Omega^2}{\p z}
\label{f_r}
\end{equation}
is an additional radial acceleration due to the radial temperature gradient and the vertical shear. Then the source terms in equation~(\ref{energy}), when averaged over a wave period, can be related to the work integral $\int f_ru_r'\,\dd t$ over the wave cycle. The additional acceleration is illustrated in Fig.~\ref{f:cartoon}, where in the local context it leads to overstability of inertial waves (i.e. the growth of body modes). 

In the second approach, we multiply equation~(\ref{Eulin2}) by $\rho u_\phi'$ to obtain an equation for the azimuthal part of the perturbation kinetic energy,
\begin{equation}
  \f{\p}{\p t}\left(\f{1}{2}\rho u_\phi'^2\right)=-\rho u_r'u_\phi'\f{1}{r}\f{\p l}{\p r}-\rho u_\phi'u_z'\f{1}{r}\f{\p l}{\p z}.
\label{energy_azimuthal}
\end{equation}
Adding this to equation~(\ref{energy_wave}) produces an equation for the total perturbation energy,
\begin{align}
  &\f{\p}{\p t}\left[\f{1}{2}\rho\left(u_r'^2+u_\phi'^2+u_z'^2\right)+\f{p'^2}{2p}\right]+\f{1}{r}\f{\p(ru_r'p')}{\p r}+\f{\p(u_z'p')}{\p z}\nonumber\\
  &\qquad=\f{\dd c_\text{s}^2}{\dd r}u_r'\rho'-\rho u_r'u_\phi'r\f{\p\Omega}{\p r}-r\rho u_z'u_\phi'r\f{\p\Omega}{\p z}.
\label{energy_reynolds}
\end{align}
In this version, the energy density (again in square brackets) includes all three components of the perturbation kinetic energy. As well as the familiar first source term involving the temperature gradient, we now have two terms resulting from the product of the Reynolds stress $-\rho u_i'u_j'$ with the shear rate. We will see in Sections~\ref{s:physical} and~\ref{s:alternative}
that the growth of travelling waves due to the VSI can be understood using either equation~(\ref{energy}) or equation~(\ref{energy_reynolds}). We will also find it instructive to compare equations (\ref{energy_azimuthal}) and (\ref{energy_reynolds}) and their source terms.

\section{Travelling waves and the VSI}
\label{s:travelling}

In this section we extend the standard theory of radially travelling waves in a thin disc, outlined in Section~\ref{s:waves}, to the case of a locally isothermal disc with an imposed radial temperature gradient. We will see that the VSI then emerges in the form of a growth of outwardly propagating inertial waves.

We relegate the formal derivation of some of the equations below to Appendix~\ref{s:asymptotic}, in order to focus here on the essential results and their physical interpretation, without the encumbrance of notation necessitated by asymptotic analysis. The interested reader can consult Appendix~\ref{s:asymptotic} for further details.

\subsection{Structure and amplitude of travelling waves}
\label{s:structure}

As indicated in Section~\ref{s:waves}, the displacement associated with a travelling wave can be written in the form
\begin{equation}
  \boldsymbol{\xi}=\text{Re}\left\{\mathcal{A}(r,t)\,\hat{\boldsymbol{\xi}}\,\exp\left[-\ii\omega t+\ii\int k(r)\,\dd r\right]\right\},
\end{equation}
where $\mathcal{A}(r,t)$ is a slowly varying (and generally complex) wave amplitude, while $\hat{\boldsymbol{\xi}}=(\hat\xi_r,\hat\xi_z)$ gives the vertical structure of the displacement associated with a particular mode. Following \citet{2000MNRAS.318.1005G}, we normalize the modes such that
\begin{equation}
  \int_{-\infty}^\infty\rho|\hat{\boldsymbol{\xi}}|^2\,\dd z=\Sigma H^2,
\end{equation}
which makes the amplitude $\mathcal{A}$ dimensionless. An explicit expression for mode~$n$, which comes from solving the ordinary differential equations governing the vertical structure of each mode, is
\begin{equation}
  \hat\xi_r=\ii X\,\text{He}_n\left(\f{z}{H}\right),\qquad
  \hat\xi_z=Z\,\text{He}_n'\left(\f{z}{H}\right),
\end{equation}
where $\text{He}_n$ is the Hermite polynomial of degree~$n$ \citep[e.g.][]{2000MNRAS.318.1005G} and
\begin{equation}
  X=\left(\f{kH\omega^2}{\omega^2-\Omega^2}\right)Z,\qquad
  Z=\left[\f{(\omega^2-\Omega^2)\Omega^2}{(\omega^4-n\Omega^4)n!}\right]^{1/2}H
\label{XZ}
\end{equation}
are two real coefficients.

We recall that $\omega$, $k$ and $n$ are connected by the dispersion relation~(\ref{dispersion}), which depends on $r$ through $\Omega$ and $c_\text{s}$. Here $\Omega$ should be interpreted as $\Omega_\text{K}(r)$, which is the leading approximation to the actual angular velocity of a thin disc. [Departures from Keplerian rotation are taken into account in the amplitude equation~(\ref{amplitude}) quoted below.] While $\omega$ and $n$ are constant for a monochromatic wave, $k$, $X$ and $Z$ depend implicitly on $r$.

With our normalization, the wave energy per unit radius is
\begin{equation}
  \mathcal{E}=\f{1}{2}\Sigma H^2\omega^2|\mathcal{A}|^2\cdot2\pi r
\label{energy_per_unit_radius}
\end{equation}
and the radial flux of wave energy is
\begin{equation}
  \mathcal{F}=\mathcal{E}v_\text{g},
\label{flux}
\end{equation}
where
\begin{equation}
  v_\text{g}=\f{\dd\omega}{\dd k}
\end{equation}
is the group velocity. A useful expression for $v_\text{g}$ can be obtained by differentiating the dispersion relation~(\ref{dispersion}):
\begin{equation}
  v_\text{g}=\f{c_\text{s}^2k}{\omega\left(1-\f{n\Omega^4}{\omega^4}\right)}.
\label{group}
\end{equation}
This is positive for outwardly propagating inertial waves ($n\ge1$, $k<0$, and $0<\omega<\Omega$).

\subsection{Growth or decay of travelling waves}

In an ideal fluid system, such as a globally isothermal or adiabatic disc, we would expect the wave energy to satisfy the conservation law
\begin{equation}
  \f{\p\mathcal{E}}{\p t}+\f{\p\mathcal{F}}{\p r}=0,
\end{equation}
meaning that it is transported at the group velocity but neither created nor destroyed. For a steady wavetrain, we would then have $\mathcal{F}=\cst$, as considered in numerous applications of wave propagation in astrophysical discs \citep[e.g.][]{1993ApJ...409..360L,1998ApJ...504..983L,2002ApJ...569..997R}.

The locally isothermal disc is non-conservative, however. The quantities $\mathcal{E}$ and $\mathcal{F}$ are closely related to those that are differentiated with respect to $t$ and $r$ in the non-conservative energy equation~(\ref{energy}). When that equation is integrated in the vertical and azimuthal directions and averaged in time over one wave period, we obtain
\begin{equation}
  \f{\p\mathcal{E}}{\p t}+\f{\p\mathcal{F}}{\p r}=\mathcal{S},
\label{energy_integrated}
\end{equation}
where the source term $\mathcal{S}=\mathcal{S}_1+\mathcal{S}_2$ consists of two parts, corresponding to the two source terms in equation~(\ref{energy}).

We can use a perturbative approach to determine the growth (or decay) of wave energy expressed by equation~\eqref{energy_integrated}. The essential assumption is that the local physics of the wave is described to a first approximation by the ideal wave theory summarized in Section~\ref{s:structure}, while the effects of the vertical shear are subdominant and cause the wave to evolve on a timescale that is long compared to the wave period. (This hierarchy is taken into account formally by the asymptotic analysis in Appendix~\ref{s:asymptotic}, and we set out the underlying assumptions more fully in Section~\ref{s:amplitude} below.) It means that the energy, flux and source term in equation~(\ref{energy_integrated}) can be estimated using the ideal wave solution. The resulting evolution is slow simply because the source term is small, as it involves the vertical shear.

The energy and flux are therefore still given by expressions (\ref{energy_per_unit_radius}) and (\ref{flux}).
After some algebra, the first source term is simply
\begin{equation}
  \mathcal{S}_1=\mathcal{F}\,\f{\dd\ln T}{\dd r},
\end{equation}
while the second term is
\begin{equation}
  \mathcal{S}_2=-\f{n\Omega^2}{\omega^2}\mathcal{S}_1.
\end{equation}
They combine to give
\begin{equation}
  \mathcal{S}=\left(1-\f{n\Omega^2}{\omega^2}\right)\mathcal{F}\,\f{\dd\ln T}{\dd r}.
\label{source_combined}
\end{equation}
The source term $\mathcal{S}_1$ is the only one operating for the density wave ($n=0$) and we will see below that this wave is modestly amplified when it propagates \emph{inwards}. For other modes, the second term $\mathcal{S}_2$ opposes the first term and is most important for inertial waves ($n\ge1$, $\omega^2<\Omega^2$), for which it exceeds $\mathcal{S}_1$ and indeed is potentially much larger.

\subsection{Physical interpretation of the growth mechanism}
\label{s:physical}

We are now in a position to develop further the physical mechanism by which an outwardly propagating inertial wave grows. 
In the context of equation~(\ref{energy}), the energy source term responsible for growth is $\mathcal{S}_2=-\rho r(\p\Omega^2/\p z)u_r'\xi_z$ and thus involves a product of the radial-velocity perturbation, the vertical displacement and the vertical shear (with a negative coefficient). As touched on qualitatively in Section 2.2.7, the growth mechanism relies on the elliptical paths of fluid elements in the meridional plane, which we have seen is predominantly clockwise above the midplane in the case of an outward inertial wave (see Fig.~\ref{f:cartoon}). Assuming that $\dd T/\dd r<0$ so that $\p\Omega/\p z<0$ above the midplane, fluid elements with a positive vertical displacement ($\xi_z>0$) have an excess of specific angular momentum relative to their surroundings and therefore experience an excess centrifugal force. If they are moving radially outwards ($u_r'>0$) then this force does positive work on the fluid element. An amplification of the wave occurs because the clockwise elliptical path naturally brings about a positive correlation between $\xi_z$ and $u_r'$. (Below the midplane the paths are predominantly anticlockwise but the vertical shear is reversed, so the amplification works in a similar way.)

The source term $\mathcal{S}_1$ is quite different in character; it involves the density perturbation $\rho'$ and relates more to the thermal physics of the perturbation than to angular momentum and the vertical shear. This source term is relatively unimportant for inertial waves, which are often well described by the anelastic approximation in which $\rho'$ is neglected. However $\mathcal{S}_1$ is the only term operating for density waves ($n=0$), in which, of course, $\rho'$ is important.

\subsection{Alternative view involving Reynolds stresses}
\label{s:alternative}

It is also illuminating to evaluate the source terms in the alternative wave energy equation~(\ref{energy_reynolds}), resulting from products of the Reynolds stress with the shear. If we substitute for $u_\phi'$ using equation~(\ref{uphip}), then there are no contributions from the products $u_r'\xi_r$ or $u_z'\xi_z$, as these are the time-derivatives of $\xi_r^2/2$ and $\xi_z^2/2$ and average to zero. The source term of interest involving the radial shear is then
\begin{equation}
  \rho u_r'\xi_z\f{\p l}{\p z}\f{\p\Omega}{\p r}
\end{equation}
and becomes (on integration and averaging, using the travelling-wave solution) $\f{3}{4}\mathcal{S}_2$,
while the source term of interest involving the vertical shear is
\begin{equation}
  \rho u_z'\xi_r\f{\p l}{\p r}\f{\p\Omega}{\p z}
\end{equation}
and becomes (similarly) $\f{1}{4}\mathcal{S}_2$.
We conclude that the source term $\mathcal{S}_2$, which is mainly responsible for the growth of outwardly propagating inertial waves, can be interpreted as issuing from Reynolds stresses acting on the radial ($\f{3}{4}$) and vertical ($\f{1}{4}$) shears. Note that the $r\phi$ Reynolds stress is much smaller than the $z\phi$ component, but makes the larger contribution in the energy equation because it is combined with the much larger radial shear. However, the $r\phi$ stress cannot lead to an accretion flow in our model, because each fluid element exactly preserves its specific angular momentum.

We now discuss the physical interpretation of the $z\phi$ and $r\phi$ Reynolds stresses. Above the midplane, the velocity perturbations $u_z'$ and $u_\phi'$ have a positive correlation in an outward inertial wave and extract energy from a vertical shear with $\p\Omega/\p z<0$. This effect can also be related to the elliptical path in Fig.~\ref{f:cartoon}, because the leading approximation to $u_\phi'$ in the wave is $-(\Omega/2)\xi_r$ and results from the radial gradient of angular momentum. In a clockwise path, $u_z'$ is negatively correlated with $\xi_r$ and therefore positively correlated with $u_\phi'$.

The $r\phi$ Reynolds stress is more subtle because $u_r'$ and $u_\phi'$ are in quadrature in the travelling wave, to a first approximation. When the vertical shear is taken into account, however, $u_\phi'$ acquires an additional contribution from $\xi_z$, as seen in equation~(\ref{uphip}). Therefore the positive correlation between $\xi_z$ and $u_r'$, discussed in the previous paragraph, brings about a positive correlation between $u_\phi'$ and $u_r'$ when $\p\Omega/\p z<0$, and therefore an $r\phi$ Reynolds stress that extracts energy from the radial shear in the Keplerian disc.

It is also illuminating to return to equation~(\ref{energy_azimuthal}) for the azimuthal perturbation kinetic energy. In a steady wavetrain, after averaging and integration, the two source terms, involving the $r\phi$ and $z\phi$ Reynolds stresses combined with the respective angular-momentum gradients, must balance each other. But the same Reynolds stresses appear as source terms in equation~(\ref{energy_reynolds}), now combined with the respective angular-velocity gradients. Since $r\f{\p\Omega}{\p z}=\f{1}{r}\f{\p l}{\p z}$, the terms involving $u_z'u_\phi'$ are the same in each equation. But since $r\f{\p\Omega}{\p r}=\f{3}{r}\f{\p l}{\p r}$ in a Keplerian disc with $\Omega\propto r^{-3/2}$ (to leading order in $H/r$, i.e.\ neglecting the vertical shear), the terms involving $u_r'u_\phi'$ are in a $3:1$ ratio in the two equations. The required balance of source terms for the azimuthal perturbation kinetic energy implies that the source term $\mathcal{S}_2$ partitions in the $3:1$ way described above.

In constructing these explanations we were reminded of the argument of \citet{1996ApJ...467...76B} that hydrodynamic turbulence could not be self-sustaining in a Keplerian disc. Their reasoning was that a positive correlation $\langle\rho u_r'u_\phi'\rangle$ is required to provide a source term in the total kinetic energy equation (\ref{energy_reynolds}) and sustain the motion against viscous dissipation, but the same correlation then provides a sink term in the azimuthal kinetic energy equation (\ref{energy_azimuthal}), meaning that the azimuthal kinetic energy cannot be sustained. In a steady outward wavetrain due to the VSI, we do find outward angular-momentum transport ($\langle\rho u_r'u_\phi'\rangle>0$); but the resulting loss of azimuthal kinetic energy is compensated by a source term from vertical shear -- an effect that is absent in the model considered by \citet{1996ApJ...467...76B}.\footnote{Their argument can be also circumvented, in principle, by means of the pressure-strain correlation, for example in strongly turbulent convection in astrophysical discs \citep{2010MNRAS.404L..64L}, although in that problem the unstable vertical stratification provides an additional source of energy.}

\subsection{Variation of the wave amplitude}
\label{s:amplitude}

\subsubsection{Wave-amplitude equation}
\label{s:amplitude_equation}

A more formal asymptotic derivation of the variation of the wave amplitude $\mathcal{A}(r,t)$ is given in Appendix~\ref{s:asymptotic}.
This approach is expected to be valid when a number of conditions are fulfilled:
\begin{enumerate}
\item the wave is of sufficiently low amplitude that the linearized equations are applicable;
\item the radial lengthscale $1/k$ associated with the wave is short compared to the radius, i.e.\ $kr\gg1$;
\item the amplitude variation occurs on a lengthscale that is long compared to $1/k$;
\item the wave forms a standing mode in the vertical direction;
\item the amplitude variation occurs on a timescale that is long compared to $1/\omega$.
\end{enumerate}
The result of this calculation is an evolution equation for the amplitude in the form
\begin{equation}
  \f{1}{v_\text{g}}\f{\p\mathcal{A}}{\p t}+\f{\p\mathcal{A}}{\p r}+\f{f}{2}\mathcal{A}=0,
\label{amplitude}
\end{equation}
where $f(r)$ is given by
\begin{equation}
  f=\f{\dd}{\dd r}\ln(\Sigma H^2rv_\text{g})-\left(1-\f{n\Omega^2}{\omega^2}\right)\f{\dd\ln T}{\dd r}.
\end{equation}
The first term in $f$ is consistent with conservation of energy, because the factor $\Sigma H^2rv_\text{g}$ appears when converting $|\mathcal{A}|^2$ into the energy flux $\mathcal{F}$. The second term, involving the temperature gradient, is non-conservative.
We will see below that equation~(\ref{amplitude}) implies the growth (in space and/or time) of outwardly propagating inertial waves, which is one way of viewing the VSI. We note that the radial temperature gradient does not affect the modal structure itself, which is determined by physics that is local in radius. The temperature gradient and the associated vertical shear appear at a higher order in the asymptotic analysis and contribute to the amplitude equation~(\ref{amplitude}).

Equation~(\ref{amplitude}) can also be related to the energy equation~(\ref{energy_integrated}). If equation~(\ref{amplitude}) is multiplied by $\Sigma H^2\omega^2\cdot2\pi rv_\text{g}\mathcal{A}^*$ and the real part is taken, then we obtain equation~(\ref{energy_integrated}) with the source term~(\ref{source_combined}). Equation~(\ref{amplitude}) contains additional information about the phase of the wave, which will be important in future work considering nonlinear interactions between waves.

We return to the five conditions listed above for the validity of this description.
Regarding condition~(i), we plan to extend equation~(\ref{amplitude}) in future work to consider (weakly) nonlinear interactions between different wave modes.
Conditions (ii) and~(iii) break down as an inertial wave approaches a Lindblad resonance at $r=R$, but is valid elsewhere. It is perhaps not obvious that conditions (iv) and~(v) break down for an inertial wave when $r\ll R$, i.e.\ when $\omega\ll\Omega$. In this limit the wave has a very short radial wavelength such that $kH\gg1$ and the vertical group velocity is sufficiently small that the wave does not form a standing mode in the time taken for the VSI to act. We will discuss the connection between this regime and the travelling-wave regime in Section~\ref{s:connecting}.

\subsubsection{Flux variation in a steady wavetrain}
\label{s:flux_steady}

Motivated by numerical simulations that exhibit extended quasi-steady trains of corrugation waves, we calculate steady wavetrain solutions to 
\eqref{amplitude}. Because disturbances propagate outward from their original radius, in order to maintain a time-independent profile, the wavetrains must emerge from an oscillatory source in the inner part of the disc. This source could take the form of persistent small-amplitude noise. 

In a steady state, equation~(\ref{amplitude}) [or equivalently equation~(\ref{energy_integrated})] implies
\begin{equation}
  \f{\dd\ln|\mathcal{F}|}{\dd r}=\left(1-\f{n\Omega^2}{\omega^2}\right)\f{\dd\ln T}{\dd r}.
\label{dfdr}
\end{equation}
If $\dd T/\dd r<0$, as we expect, then the non-conservative term makes $|\mathcal{F}|$ increase with $r$ for inertial waves ($n\ge1$, $\omega^2<\Omega^2$) and decrease with $r$ for acoustic waves ($n\ge1$, $\omega^2>n\Omega^2$) or for the density wave ($n=0$). Amplification of the flux, following the direction of propagation of the wave, therefore occurs for outgoing inertial waves, ingoing acoustic waves and ingoing density waves. At least in the case of the inertial waves, we can say that the VSI is acting as a `convective' instability, such that the wave amplitude increases following its (outward) propagation, but not at a fixed point in space.

\begin{figure}
    \centering
    \includegraphics[width=8cm]{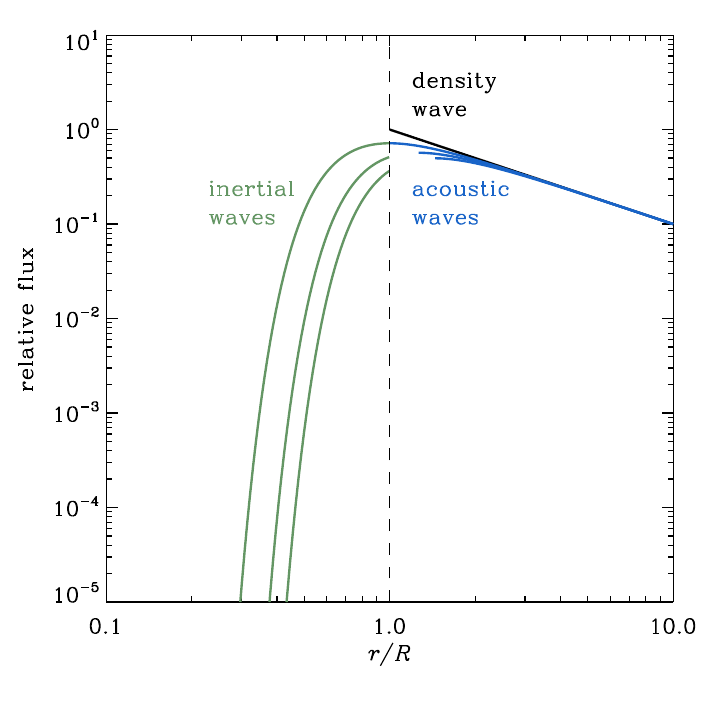}
    \caption{Relative variation of the radial flux of wave energy with radius (equation~\ref{fq}) for waves with vertical mode numbers $n=0$ (black line: density wave), $1$, $2$ and $3$ (coloured curves, from top to bottom: inertial and acoustic waves) in a Keplerian disc with $q=-1$, i.e.\ $c_\text{s}^2\propto T\propto r^{-1}$. The Lindblad resonance at $r=R$ is indicated by a vertical dashed line. The $n=0$ density wave propagates only where $r>R$. The $n=1$ wave propagates at all radii. For waves with $n>1$, the inertial and acoustic branches are separated by a forbidden region. If the value of $q$ is varied, then the (logarithmic) vertical axis of this figure scales with $q$, i.e.\ $\ln|\mathcal{F}|\propto q$.}
    \label{f:flux}
\end{figure}

If $T\propto r^q$, where $q=\dd\ln T/\dd\ln r$ is a constant (presumably negative), then equation~(\ref{dfdr}) can be integrated to give
\begin{equation}
  |\mathcal{F}|\propto c_\text{s}^2\exp\left(\f{nq\Omega^2}{3\omega^2}\right)=c_\text{s}^2\exp\left(\f{nqR^3}{3r^3}\right),
\label{fq}
\end{equation}
where we recall that $R$ is the radius of the Lindblad resonance at which $\omega=\Omega$. The radial variation of flux with $r$ for different modes is illustrated in Fig.~\ref{f:flux}, which shows that the largest growth occurs for $r\ll R$. The effect increases with $n$ and $|q|$.

The behaviour $|\mathcal{F}|\propto c_\text{s}^2$ for density waves ($n=0$) in locally isothermal discs was highlighted by \citet[][equation 46]{2020ApJ...892...65M}, who quote \citet{2016ApJ...832..166L} as giving the result without proof. It implies that ingoing density waves are amplified while outgoing waves are damped. The effect is modest because the exponential factor is absent for the $n=0$ mode, but it has been recognized as being important in the interpretation of simulations of planet--disc interactions.

We obtain here a generalization and extension of their result. Modes with $n\ge1$ experience an additional variation of $|\mathcal{F}|$ with $r$ because of the exponential factor in equation~(\ref{fq}). For acoustic waves this produces an effect that is similar to that for the density waves, but flattens out as the turning point at which $\omega=\sqrt{n}\,\Omega$ is approached. For inertial waves it implies the opposite behaviour: the flux increases with radius, potentially by many orders of magnitude, and this can be seen as a manifestation of the VSI.

In principle, if no nonlinear effects intervened, an inertial wave with $n\ge2$ propagating outwards from the inner part of the disc would grow until it reaches the Lindblad resonance, where it would reflect and propagate inwards, decaying as it does so. This would set up a standing wave in the radial direction, with an amplitude that depends strongly on radius. The $n=1$ mode, however, would propagate through the Lindblad resonance without reflection, and convert to an acoustic wave. In practice, simulations show that the wave transfers energy to one of lower frequency (and so larger $R$), and the process repeats cyclically \citep{2022MNRAS.514.4581S}.

The energy budget for a steady outward inertial wavetrain involves an extraction of energy from the external radiation field at all radii $r<R$, allowing $|\mathcal{F}|$ to grow with $r$. In principle, only a tiny seed is required at some $r\ll R$.

\subsubsection{Variation of the vertical displacement}

A useful dimensionless measure of the amplitude of the wave is $(Z/H)\mathcal{A}$, which is proportional to the vertical displacement in units of the local scaleheight. In particular, for the $n=1$ `corrugation' mode that has a vertical displacement independent of $z$, $(Z/H)\mathcal{A}$ is precisely (apart from the phase factor depending on $r$ and $t$) the vertical displacement in units of the scaleheight.

\begin{figure}
    \centering
    \includegraphics[width=8cm]{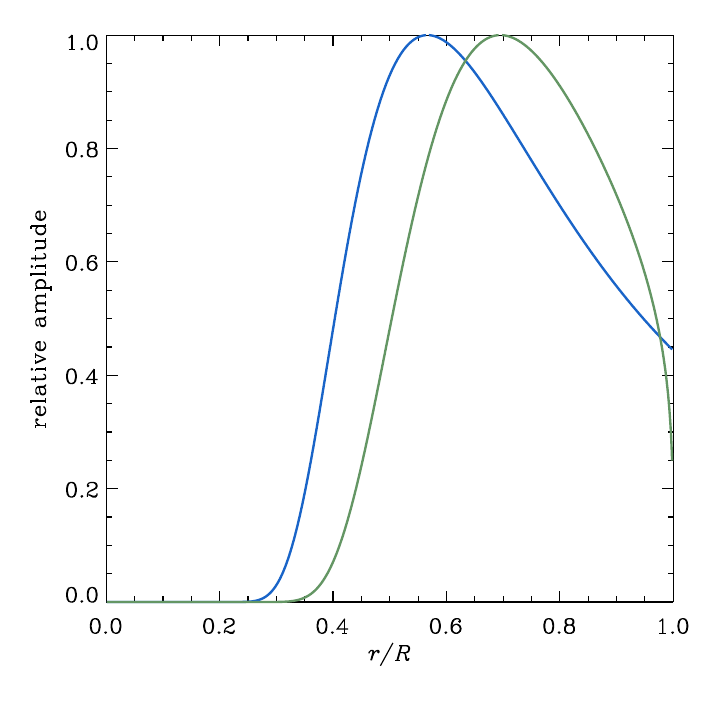}
    \caption{Relative variation of the amplitude $(Z/H)|\mathcal{A}|$ (a dimensionless measure of the vertical displacement) with radius (equation~\ref{za}) for waves with vertical mode numbers $n=1$ (blue curve) and $n=2$ (green curve), in a disc with $\Sigma\propto r^{-1}$ and $T\propto r^{-1}$. (The extremely low amplitudes plotted for small radii should not be trusted owing to the breakdown of the underlying assumptions as discussed in Section~\ref{s:amplitude_equation}.)}
    \label{f:amplitude}
\end{figure}

Using equations~(\ref{XZ}), (\ref{flux}) and (\ref{group}), we can relate this quantity to the flux by
\begin{equation}
  \left|\f{Z}{H}\mathcal{A}\right|^2=\f{(\omega^2-\Omega^2)}{k\omega^5}\f{\mathcal{F}}{n!\,\pi r\Sigma H^4}.
\end{equation}
Using the dispersion relation, this can also be written as
\begin{equation}
  \left|\f{Z}{H}\mathcal{A}\right|^2=\left(\f{\Omega^2-\omega^2}{n\Omega^2-\omega^2}\right)^{1/2}\f{c_\text{s}}{\omega^4}\f{|\mathcal{F}|}{n!\,\pi r\Sigma H^4}.
\end{equation}
Under the assumption leading to equation~(\ref{fq}), we then find
\begin{equation}
  \left|\f{Z}{H}\mathcal{A}\right|\propto\left(\f{1-\f{r^3}{R^3}}{n-\f{r^3}{R^3}}\right)^{1/4}\left[\f{1}{\Sigma r^{7+(q/2)}}\right]^{1/2}\exp\left(\f{nqR^3}{6r^3}\right).
\label{za}
\end{equation}
Even for $n=1$, this expression implies powerful growth of the amplitude at small radii, up to $r/R$ of around $0.6$, depending on the profiles of $c_\text{s}$ and $\Sigma$. This variation is illustrated in Fig.~\ref{f:amplitude}. Note that, for $a,b>0$, $r^{-a}\exp(-bR^3/3r^3)$ peaks at $r/R=(b/a)^{1/3}$, which has quite a weak dependence on $a$ and $b$. The peak occurs because of a competition between the growth of the amplitude due to the VSI (accounted for mainly by the exponential function) and the decline of the amplitude due to geometrical dilution as the wave propagates outwards into a much larger volume (which accounts for most of the negative powers of $r$).

\subsubsection{Local growth}

If the group propagation is ignored, equation~(\ref{amplitude}) suggests a local growth rate of
\begin{equation}
  \f{v_\text{g}}{2}\left(1-\f{n\Omega^2}{\omega^2}\right)\f{\dd\ln T}{\dd r},
\end{equation}
which can also be expressed as
\begin{equation}
  -\f{q}{2}\f{H}{r}\Omega\left[\f{(n\Omega^2-\omega^2)^{3/2}(\Omega^2-\omega^2)^{1/2}}{n\Omega^4-\omega^4}\right]
\label{local_growth_rate}
\end{equation}
and simplifies to
\begin{equation}
  -\f{q}{2}\f{H}{r}\Omega\left(\f{\Omega^2-\omega^2}{\Omega^2+\omega^2}\right)
\end{equation}
in the case $n=1$.

Far inside the Lindblad resonance ($\omega^2\ll\Omega^2$, i.e.\ $r^3\ll R^3$), the local growth rate (\ref{local_growth_rate}) approximates to
\begin{equation}
  -\f{q}{2}\f{H}{r}\Omega\sqrt{n}.
\end{equation}
This expression agrees with equation~36 of \citet{2015MNRAS.450...21B}, who noted that it ceases to be valid if $n$ is too large. The limit in which they obtained this result was expressed by them as $|kq|\ll1$, but owing to the scaling of the variables that they employed, this corresponds in our notation to $|kq|(H/r)^2\ll1$. Assuming $|q|=O(1)$, this inequality is compatible with the assumptions we have made here, provided that $(r/H)\gg|kH|\gg1$, which covers a reasonable range of radial wavenumbers and locations in the case of a thin disc. We will discuss the connection with the low-frequency regime in Section~\ref{s:connecting}.

Finally, the growth rate can be compared with the `propagation rate' $v_\text{g}/r$, i.e.\ $\dd\ln r/\dd t$ following the group propagation. 
If the former supersedes the latter, the disturbance is likely to grow and saturate in place through nonlinear effects.
If the reverse holds, then the disturbance will travel significant distances before it grows appreciably.
The ratio of the local growth rate to the global propagation rate is
\begin{equation}
  \f{|q|}{2}\left(\f{n\Omega^2}{\omega^2}-1\right),
\end{equation}
which agrees with $\dd\ln\sqrt{|\mathcal{F}|}/\dd\ln r$ for a steady wavetrain.
The comparison suggests that, for propagation to be more important than local growth and for the wave to be inside its Lindblad resonance, we require
\begin{equation}
  \f{n}{1+\f{2}{|q|}}\lesssim\left(\f{r}{R}\right)^3\lesssim1.
\end{equation}
Such a region exists only if $n<1+\f{2}{|q|}$, i.e.\ only for $n=1$ or $n=2$ in the case $q=-1$. Therefore the `convective', travelling nature of the VSI is likely to be observed only for these two wave modes. This result explains the importance of the `corrugation' and `breathing' modes in the large-scale development of the VSI in numerical simulations.

\subsubsection{An evolving wavetrain}

So far, we have mainly examined steady wavetrain solutions, which require a continual supply of (small-amplitude) forcing in the inner disc (which may be modelled as a constant wave flux at the inner boundary). However, it is possible to construct time-dependent solutions for arbitrary initial conditions with or without forcing, which may be useful in the interpretation of numerical simulations. We give the details in Appendix~\ref{s:evolving}. There we find that any initial condition propagates radially outwards. At sufficiently long times, the disc is then dominated by the steady wavetrain issuing from the forcing at the inner boundary.

\section{Connecting the inner and outer regions}
\label{s:connecting}

In the preceding section, we demonstrated that the VSI body modes can either grow essentially in place and presumably develop into a nonlinear regime, or can travel global distances before growing appreciably. The former 
behaviour is exhibited by all linear modes with $n>2$ at all radii and for those with $n=1$ or $n=2$ at sufficiently small radii. This result suggests that protoplanetary discs possess an `outer' region of growing travelling waves and an `inner' region of locally growing disturbances, although this does assume that the waves of interest have the same or similar frequencies. In this section we examine in more detail how these two regions are able to connect. 

To better develop the analysis, we employ an alternative formalism to that used so far; this is outlined in Appendix~\ref{s:hermite}. There we project the full linearized equations onto a basis of Hermite polynomials in the vertical direction. This is a widely used technique for waves in vertically isothermal discs \citep[e.g.][]{1987PASJ...39..457O,2006MNRAS.368..917Z,2008MNRAS.388.1372O}. We argue that the coupling between different modes is weak and that the local dispersion for mode~$n$ in the short-wavelength limit is
\begin{equation}
  \left(\f{\omega^2}{\Omega^2}-1\right)\left(\f{\omega^2}{\Omega^2}-n\right)+\ii nq\epsilon kH=\f{\omega^2}{\Omega^2}(kH)^2,
\label{ldr_hermite}
\end{equation}
where $\epsilon=H/r$.

If $n$, $q$ and $kH$ are regarded as quantities of order unity, while $\epsilon\ll1$ for a thin disc, then it would be justified to neglect the term $\ii nq\epsilon kH$ in equation~(\ref{ldr_hermite}). It would then agree with the usual local dispersion relation~(\ref{dispersion}) for the isothermal disc and we would recover the wave modes described in Section~\ref{s:waves}, uninfluenced at this order by the radial temperature gradient and the VSI. If the analysis were carried to the next order, to determine the radial variation of the wave amplitude, we would recover the results of Section~\ref{s:amplitude} in which the VSI appears as a growth of the wave amplitude over many wavelengths (or wave periods).

On the other hand, if $kH$ and/or $n$ are large such that the term $\ii nq\epsilon kH$ cannot be neglected, then the VSI appears within the local dispersion relation, which then becomes complex. In the low-frequency limit $\omega^2\ll\Omega^2$ this gives (for $n\ge1$)
\begin{equation}
  \f{\omega^2}{\Omega^2}\approx\f{n\left(1+\ii q\epsilon kH\right)}{(kH)^2},
\label{ldr_bl15}
\end{equation}
in agreement with equation~34 of \citet{2015MNRAS.450...21B}.\footnote{Interestingly, \citet{2015MNRAS.450...21B} obtained their dispersion relation by considering solutions described by Hermite polynomials of complex argument, whereas we have obtained the same result using real polynomials.}

The dispersion relation~(\ref{ldr_bl15}) can be read in (at least) two different ways. If $k$ is regarded as real, which is equivalent to assuming that the disturbance has an oscillatory spatial structure in $r$ (as well as in $z$ because of the Hermite polynomial), then a complex frequency $\omega$ is obtained. The VSI then appears as a temporal growth of the disturbance; the growth rate $\text{Im}(\omega)$ tends to $\f{1}{2}|q|\epsilon\Omega$ in the limit $|q\epsilon kH|\ll1$ and is reduced for larger $k$.

Alternatively, if $\omega$ is regarded as real, which is appropriate for a wave driven by an oscillatory source, then a complex radial wavenumber $k$ is obtained. In fact, equation~(\ref{ldr_bl15}) is then a quadratic equation with real coefficients for the quantity $\ii kH$:
\begin{equation}
  \f{\omega^2}{n\Omega^2}(\ii kH)^2+q\epsilon(\ii kH)+1=0.
\end{equation}
The discriminant of the quadratic equation is
\begin{equation}
   \Delta=(q\epsilon)^2-\f{4\omega^2}{n\Omega^2}.
\end{equation}
If $\Delta>0$, then both roots $\ii kH$ are real and the waves are evanescent in the radial direction. However, if $\Delta<0$, then the roots are complex and so are the values of $k$:
\begin{equation}
  kH={\f{n\Omega^2}{2\omega^2}}\left(\ii q\epsilon\pm\sqrt{|\Delta|}\right),
\label{k_complex}
\end{equation}
implying
\begin{equation}
  \text{Im}(k)={\f{n\Omega^2}{2\omega^2}}\f{q}{r}.
\end{equation}
If $q=\cst$ then the non-oscillatory spatial part of the wave solution $\exp(\ii\int k\,\dd r)$ is
\begin{equation}
  \exp\left[-\int\text{Im}(k)\,\dd r\right]\propto\exp\left({\f{nqR^3}{6r^3}}\right),
\end{equation}
which accounts for the real exponential factor in the variation of amplitude (or flux, when squared) found in Section~\ref{s:flux_steady}.

The latter viewpoint confirms the picture that outwardly travelling inertial waves grow in space because of the VSI. The condition $\Delta<0$ also implies that the waves travel when
\begin{equation}
  \f{\omega}{\Omega}=\left(\f{r}{R}\right)^{3/2}>\f{|q|}{2}\f{H}{r}\sqrt{n},
\end{equation}
i.e.\ outside a critical radius that is typically much smaller than that of the Lindblad resonance. For example, in the case $n=1$, $H/r=0.05$, $|q|=1$, the critical radius is $r_1=0.0855\,R$.

The transition between non-propagating disturbances that grow in place and propagating disturbances that grow as they travel has an interesting structure that may merit further investigation. This can be seen by considering the group velocity $v_\text{g}=\dd\omega/\dd k$ derived from the dispersion relation~(\ref{ldr_bl15}) in the case of a real frequency~$\omega$. We find
\begin{equation}
  \text{Re}\left(v_\text{g}\right)=\pm\f{H\omega^2}{\Omega\sqrt{n}}\left(1-\f{1}{2}nq^2\epsilon^2\f{\Omega^2}{\omega^2}\right)\left(1-\f{1}{4}nq^2\epsilon^2\f{\Omega^2}{\omega^2}\right)^{1/2}.
\end{equation}
The two bracketed factors represent corrections to the group velocity of a low-frequency inertial wave due to the radial temperature gradient and associated vertical shear. The second bracketed factor vanishes at the critical radius $r_1$, where $\Delta=0$. The first bracketed factor is negative here and only becomes positive beyond a radius $r_2$ that is slightly larger than the critical radius ($r_2/r_1=2^{1/3}\approx1.26$ in the case $\epsilon=\cst$). This implies that the waves actually propagate inwards in the region $r_1<r<r_2$, so the outwardly propagating waves for large $r$ can be seen as originating around $r=r_2$.

\section{Implications}
\label{s:implications}

We now step back from the detailed analysis and consider the application of our findings to the VSI in both numerical simulations and protoplanetary discs.

We have emphasized the `convective' nature of the VSI, which can take the form of an outwardly propagating inertial wave. In a typical disc model such as one with $T\propto r^{-1}$, this description may be relevant only for the $n=1$ (corrugation) and $n=2$ (breathing) modes. For these modes, radial propagation at the group velocity dominates over growth in place for a broad range of radii interior to the wave's Lindblad resonance. In principle, this provides a linear mechanism for the VSI to reach a finite amplitude and to produce a steady wavetrain over this range of radii. Numerical simulations \citep[e.g.][]{2022MNRAS.514.4581S} suggest that this process is repeated cyclically, with each wave handing over, presumably through weakly nonlinear mode coupling, to another wave of lower frequency that propagates further out.

In the early stages of a numerical simulation, there is much activity in rapidly growing surface modes and body modes with $n>2$. These presumably grow mostly in place and break rapidly when they reach nonlinear amplitudes. Then the body modes with $n=2$ and $n=1$ can emerge to dominate the scene. We are not aware of any explanation for why the slower-growing $n=1$ mode supplants the $n=2$ mode in numerical simulations.

Important questions remain about the source of the waves. The outwardly travelling nature of the VSI means that the wave activity at a certain location in the disc is strongly influenced by what is happening at smaller radii. In the quasi-steady final state found in numerical simulations, it appears that the first wave to emerge from the inner region has a frequency that is a small fraction of the Keplerian frequency at the inner boundary. For this wave, the innermost part of the computational domain is in the low-frequency regime in which radial propagation is unimportant and the VSI causes a growth of disturbances in place. Although we have discussed the transition between this regime and that of travelling waves, we have not resolved the question of how the frequency of the emerging wavetrain is selected. This may depend on nonlinear behaviour, because the VSI cannot saturate by a linear mechanism in the low-frequency regime. Uncomfortably, it may also depend on computational details to do with the handling of the inner boundary and any wave-damping zones that are introduced to quell numerical instabilities.

In an actual protoplanetary disc there are likely to be multiple sources of fluctuations resulting from the complicated physics of weakly ionized dusty gas with magnetic fields, outflows, zonal flows, vortices, planetesimals, planetary embryos, etc. In principle, each such source could give rise to a train of inertial waves that grows through the VSI as it travels outwards. Or, as in a numerical simulation, the innermost part of the disc may provide the dominant source of activity. Identifying the most important of these sources would be critical to understanding the structure of the wavetrains that emerge.

The wavetrains are likely to be subject to a number of secondary processes. As seen in high-resolution numerical simulations, the shear involved in the waves may give rise to vortices in either the $rz$-plane or the $r\phi$-plane. Wave breaking or shocking might occur at sufficiently high altitudes. Nonlinear mode couplings may also disrupt the orderly process described in this paper.

\section{Conclusion}
\label{s:conclusion}

We have revisited the linear theory of the VSI in protoplanetary discs with an imposed radial temperature gradient. Our analysis is most relevant to the `body modes' known as `corrugation' and `breathing' modes, which fall into the category of inertial waves. We have developed a quantitative description of the spatial structure of these waves in a standard protoplanetary disc model and the evolution of their amplitudes in space and time. As found in some numerical simulations, these modes tend to take the form of a train of waves that propagates outwards towards a Lindblad resonance. The radial propagation dominates over the growth mechanism for a broad range of radii, meaning that the waves develop more in space than in time.

We have made a detailed analysis of the mechanism by which the radial temperature gradient and associated vertical shear causes the waves to grow. More energy is in fact extracted by the inertial waves from the radial (Keplerian) shear than from the vertical shear. We have also explained the relationship between the VSI and the non-conservative behaviour of density waves in locally isothermal discs.

Although we have also discussed the connection between the travelling-wave regime of the VSI and the lower-frequency regime in which the disturbances grow in place, our analysis does not explain the selection of wave frequencies observed in numerical simulations. This, and the (presumably nonlinear) transfer from one wavetrain to another, remain to be investigated in future work.

\section*{Acknowlegements}

We are grateful to the referee for a thorough review of the original manuscript that enabled several improvements to be made. This research was supported by STFC through grant ST/X001113/1.

\section*{Data Availability}

The data underlying this article will be shared on reasonable request to the corresponding author.

\bibliographystyle{mnras}
\bibliography{main}

\appendix

\section{Thermal energy in a locally isothermal disc}
\label{s:energy}

According to equation~(\ref{energy_mechanical}), the source term for mechanical energy, per unit volume, is $p\grad\bcdot\vecu=p\,\DD\ln v/\DD t$, where $v=1/\rho$ is the specific volume, so the energy source per unit mass is $p\,\DD v/\DD t$. In general, the differential $p\,\dd v$ can be expressed as $-\dd e+T\,\dd s$, where $e$ is the specific internal energy, $T$ is the temperature and $s$ is the specific entropy. If the gas were to behave adiabatically, such that $\DD s/\DD t=0$, then $p\,\DD v/\DD t$ could be rewritten as $-\DD e/\DD t$ and $e$ could then be included in the density and flux terms on the left-hand side to make a total energy equation in strictly conservative form, i.e.\ with no source term. Alternatively, if the gas were globally isothermal, with a uniform temperature $T$ and sound speed $c_\text{s}$, then $p\,\dd v$ could be rewritten as $-\dd f$, where $f=e-Ts$ is the specific Helmholtz free energy, and $f$ could then be included in the density and flux terms on the left-hand side to make a total energy equation in strictly conservative form.\footnote{For a perfect gas, $e=RT/(\gamma-1)+\cst$, where $RT=p/\rho=c_\text{s}^2$, while $f=RT(1-\ln T)/(\gamma-1)+RT\ln \rho+\cst$. Under globally isothermal conditions, the relevant free energy is therefore $c_\text{s}^2\ln\rho+\cst$.} But under the locally isothermal conditions considered in this paper, $p\,\dd v$ is not an exact differential, there is no strictly conserved form of energy and the energy equation requires a source term. The source term for mechanical energy, per unit mass, can be written as $-c_\text{s}^2\,\DD\ln\rho/\DD t=-\DD(c_\text{s}^2\ln\rho)/\DD t+(\ln\rho)\DD c_\text{s}^2/\DD t$ and the first part incorporated with the mechanical energy to make equation~(\ref{energy_non-conservation}).

\section{Asymptotic analysis of travelling VSI waves}
\label{s:asymptotic}

In this appendix we give a detailed derivation of the properties of waves in a thin disc and the growth of inertial waves as they travel outwards. The analysis is based on an asymptotic expansion of the linearized equations in powers of the aspect ratio of the disc. This approach turns out to be useful because it formalizes the separation of scales that occurs naturally in the physical problem. The dispersion relation implies that the radial wavelength is typically comparable to, or even shorter than, the vertical scaleheight of the disc, and therefore much smaller that the radial coordinate, which is the scale on which the properties of the disc change in the radial direction. Our approach is related to other short-wavelength asymptotic analyses such as the WKB approximation.

As far as the VSI is concerned, the vertical shear is smaller than the the radial shear (or orbital frequency) by a factor comparable to the aspect ratio. This weak baroclinicity means that the growth mechanism is subdominant to the basic physics of the travelling inertial wave and therefore appears at a higher order in the asymptotic analysis. Nevertheless, a very large growth factor can be accumulated when the wave travels a distance such that $r$ increases by a factor of a few, because of the asymptotically large number of wavelengths that are executed.

Even in the case of a steady wavetrain, this behaviour can also be related to a separation of timescales in the problem. The radial group velocity is slow (less than the sound speed, which is much smaller than the orbital speed) and the temporal growth rate of the VSI is also small compared to the orbital frequency, but in such a way that a large growth factor can be accumulated when the wave travels over a factor of a few in $r$.

\subsection{Thin-disc asymptotics}

To help order the expansion we introduce a small parameter $\epsilon\ll1$ representing the thinness of the disc, such that $H/r=O(\epsilon)$. We introduce stretched vertical and time variables $\zeta=z/\epsilon$ and $\tau=\epsilon t$ to resolve the rapid variation of fluid properties in the vertical direction and to allow for a slow evolution in time. We also define a rescaled sound speed $c_0=c_\text{s}/\epsilon$.

The gravitational potential $\Phi=-GM/\sqrt{r^2+z^2}$ of the central mass can be expanded in a Taylor series about the midplane of the thin disc:
\begin{equation}
  \Phi=-\f{GM}{r}+\epsilon^2\f{GM}{2r^3}\zeta^2+O(\epsilon^4).
\end{equation}

\subsection{Basic state}

In this approximation, the basic state satisfying the equilibrium balances in Section~\ref{s:equilibrium} can be expanded as
\begin{align}
  &\rho=\rho_0(r,\zeta)+\epsilon^2\rho_2(r,\zeta)+O(\epsilon^4), \\
  &\Omega=\Omega_0(r)+\epsilon^2\Omega_2(r,\zeta)+O(\epsilon^4),
\end{align}
where $\Omega_0=\sqrt{GM/r^3}$ is the Keplerian angular velocity. After cancelling the terms involved in the Keplerian balance, 
equations \eqref{eqm1} and \eqref{eqm2} at leading order become
\begin{align}
  -2r\Omega_0\Omega_2&=\f{3GM}{2r^4}\zeta^2-\f{1}{\rho_0}\f{\p(c_0^2\rho_0)}{\p r},\\
  0&=-\f{GM}{r^3}\zeta-c^2\f{\p\ln\rho_0}{\p\zeta},
\end{align}
and their solution is
\begin{align}
  &\rho_0=\rho_{\text{m}0}\exp\left(-\f{\zeta^2}{2H_0^2}\right),\qquad
  H_0=\f{c_0}{\Omega_0},\\
  &\Omega_2=\frac{1}{2r\Omega_0}\left[c_0^2\f{\dd\ln\rho_{\text{m}0}}{\dd r}+\left(1+\f{\zeta^2}{2H^2}\right)\f{\dd c_0^2}{\dd r}\right],\label{omega2}
\end{align}
where $\rho_{\text{m}0}(r)$ and $c_0(r)$ are the radially varying midplane density and (scaled) sound speed, both of which are prescribed at the outset. This solution is equivalent to equations (\ref{rho_thin}) and (\ref{omega_thin}) and can be shown to agree with equations 12--13 of Nelson et al.\ (2013) when they are expanded to the same order in $\epsilon$. Of greatest importance is the vertical variation of $\Omega$ associated with the radial variation of sound speed, as seen in the term $\propto\zeta^2$ in the expression~(\ref{omega2}) for $\Omega_2$.

\subsection{Linear perturbations}
\label{s:linpert}

We now examine perturbations to this equilibrium state. As described in Section~\ref{s:waves}, the waves of interest have a rapid variation of phase in radius and time as well as an amplitude that evolves slowly in radius and time. They also have a non-trivial modal structure in the vertical direction. The asymptotic expression of a perturbation variable such as $\xi_r$ is
\begin{equation}
  \xi_r=\epsilon\,\text{Re}\left\{E\left[\xi_{r0}(r,\zeta,\tau)+\epsilon\xi_{r1}(r,\zeta,\tau)+\cdots\right]\right\},
\end{equation}
where
\begin{equation}
  E=\exp\left[-\ii\omega_0t+\f{\ii}{\epsilon}\int k(r)\,\dd r\right]
\end{equation}
is a rapidly varying phase factor, $\omega_0$ is the frequency with respect to the fast time variable and $k(r)$ is the local radial wavenumber. At leading order, this expression is equivalent to equation~(\ref{wkb}), with $\xi_{r0}$ corresponding to the leading-order wave amplitude $X'$. The asymptotic expansion continues with higher-order corrections $\xi_{r1}$, etc. Similar expansions apply for $\xi_z$, $\rho'$ and $p'$, although the leading term in $\rho'$ scales with $\epsilon^0$ and that in $p'$ scales with~$\epsilon^2$.

When a radial derivative acts on $\xi_r$ (for example), the largest effect comes from the phase factor $E$ and results in a multiplication by $\ii k/\epsilon$. At the next order in $\epsilon$, the radial derivative acts on the amplitude function(s). Similarly, a time derivative produces a multiplication by $-\ii\omega_0$ plus a subdominant effect involving a derivative of the amplitude with respect to the slow time variable $\tau$.

\subsection{Vertical profiles and dispersion relation}
\label{s:leading}

We substitute the form of the perturbations assumed in Section~\ref{s:linpert} into equations (\ref{d2xir}) and (\ref{d2xiz}), along with (\ref{rhop}) and (\ref{pp}), and collect terms at each order in $\epsilon$.  At the leading order we obtain
\begin{align}
  &\left(-\omega_0^2+\Omega_0^2\right)\xi_{r0}=-\ii k\left(\f{p_0'}{\rho_0}\right),\\
  &-\omega_0^2\xi_{z0}=-\f{\p}{\p\zeta}\left(\f{p_0'}{\rho_0}\right),\\
  &\f{p_0'}{\rho_0}=\f{c_0^2\rho_0'}{\rho_0}=-c_0^2\left(\ii k\xi_{r0}+\f{\p\xi_{z0}}{\p\zeta}\right)+\Omega_0^2\zeta\xi_{z0}.
\end{align}
These equations are the same as those that would be obtained in the local approximation for wave solutions $\propto\exp(\ii kx-\ii\omega_0t)$ and are equivalent to those solved by \citet{1993ApJ...409..360L} in the case $\gamma=1$. We eliminate $p_0'$ and simplify to find
\begin{align}
  &\left(\Omega_0^2-\omega_0^2+c_0^2k^2\right)\xi_{r0}+\ii k\left(\Omega_0^2\zeta\xi_{z0}-c_0^2\f{\p\xi_{z0}}{\p\zeta}\right)=0,\label{xir0}\\
  &\left(\Omega_0^2-\omega_0^2\right)\xi_{z0}+\Omega_0^2\zeta\f{\p\xi_{z0}}{\p\zeta}-c_0^2\f{\p}{\p\zeta}\left(\ii k\xi_{r0}+\f{\p\xi_{z0}}{\p\zeta}\right)=0.\label{xiz0}
\end{align}
This is a second-order system of ordinary differential equations (ODEs) in $\zeta$ at each $r$ (and $\tau$). As is well known \citep{1987PASJ...39..457O}, the relevant solutions are Hermite polynomials:
\begin{equation}
  \xi_{r0}=A\,\text{He}_n\left(\f{\zeta}{H_0}\right),\qquad
  \xi_{z0}=B\,\text{He}_n'\left(\f{\zeta}{H_0}\right).
\end{equation}
Here $A$ and $B$ are constants as far as the ODE system in $\zeta$ is concerned, although in the global solution they depend on $r$ (and $\tau$) in a way that remains to be determined.
Using the Hermite differential equation and its derivative,
\begin{align}
  &x\,\text{He}_n'(x)-\text{He}_n''(x)=n\,\text{He}_n(x),\\
  &x\,\text{He}_n''(x)-\text{He}_n'''(x)=(n-1)\,\text{He}_n'(x),
\end{align}
we see that equations (\ref{xir0})--(\ref{xiz0}) are satisfied provided that
\begin{align}
  &\left(\Omega_0^2-\omega_0^2+c_0^2k^2\right)A+\ii k\Omega_0^2H_0nB=0,\\
  &\left(n\Omega_0^2-\omega_0^2\right)B-\f{\ii kc_0^2}{H_0}A=0,
\end{align}
leading to the dispersion relation
\begin{equation}
  \left(\Omega_0^2-\omega_0^2+c_0^2k^2\right)\left(n\Omega_0^2-\omega_0^2\right)=n\Omega_0^2c_0^2k^2,
\end{equation}
or, equivalently,
\begin{equation}
  \left(\Omega_0^2-\omega_0^2\right)\left(n\Omega_0^2-\omega_0^2\right)=(\omega_0c_0k)^2,
\end{equation}
in agreement with equation~(\ref{dispersion}) quoted in Section~2. The ratio of $A$ and $B$ is also determined by the above algebraic equations, so that what remains is to determine the variation of the amplitude $A$ with $r$ (and $\tau$).

It will be helpful for interpretation to replace $A$ and $B$ with a common, normalized amplitude $\mathcal{A}$. We therefore write the leading-order solution in the form
\begin{equation}
  \begin{pmatrix}\xi_r\\\xi_z\end{pmatrix}=\mathcal{A}(r,t)\begin{pmatrix}\hat\xi_r\\\hat\xi_z\end{pmatrix},
\end{equation}
where $\mathcal{A}(r,\tau)$ is a slowly varying (dimensionless, complex, normalized) wave amplitude and $(\hat\xi_r,\hat\xi_z)^\text{T}$ is a normalized modal displacement defined by
\begin{equation}
  \hat\xi_r=\ii X\,\text{He}_n\left(\f{\zeta}{H_0}\right),\qquad
  \hat\xi_z=Z\,\text{He}_n'\left(\f{\zeta}{H_0}\right),
\end{equation}
with
\begin{equation}
  X=\left(\f{kH_0\omega^2}{\omega^2-\Omega^2}\right)Z,\qquad
  Z=\left[\f{(\omega^2-\Omega^2)\Omega^2}{(\omega^4-n\Omega^4)n!}\right]^{1/2}H_0.
\end{equation}
This is normalized as in \citet{2000MNRAS.318.1005G} such that
\begin{equation}
  \int_{-\infty}^\infty\rho_0\left(|\hat\xi_r|^2+|\hat\xi_z|^2\right)\,\dd\zeta=\Sigma_0H_0^2,
\end{equation}
where $\Sigma_0=\int\rho_0\,\dd\zeta$.

\subsection{Regions of propagation}

In an unbounded Keplerian disc, an axisymmetric wave of any frequency $\omega_0>0$ has a unique Lindblad resonance radius $R$ at which $\Omega_0=\omega_0$. A density wave ($n=0$) is able to propagate (i.e.\ it has $k^2>0$) where $\omega_0>\Omega_0$, i.e.\ where $r>R$. The Lindblad resonance acts as a turning point, reflecting an ingoing density wave to produce an outgoing wave. An inertial wave with $n>1$ is able to propagate where $\omega_0<\Omega_0$, i.e.\ where $r<R$. The Lindblad resonance again acts as a turning point, reflecting an outgoing inertial wave to produce an ingoing wave. An acoustic wave with $n>1$ is confined to the region $r>n^{1/3}R$ in which $\omega_0^2>n\Omega_0^2$. The $n=1$ mode has a special behaviour in a Keplerian disc. It can propagate as an inertial wave where $r<R$ and as an acoustic wave where $r>R$, passing smoothly through the hybrid Lindblad/vertical resonance at $r=R$. This hybrid resonance was analysed and simulated by \citet{2002MNRAS.332..575B}.

\subsection{Amplitude equation and non-conservation of wave action}

If we carry the calculations of Section~\ref{s:leading} to the next order in $\epsilon$, we obtain
\begin{align}
  &\left(\Omega_0^2-\omega_0^2+c_0^2k^2\right)\xi_{r1}+\ii k\left(\Omega_0^2\zeta\xi_{z1}-c_0^2\f{\p\xi_{z1}}{\p\zeta}\right)=F_r,\label{xir1}\\
  &\left(\Omega_0^2-\omega_0^2\right)\xi_{z1}+\Omega_0^2\zeta\f{\p\xi_{z1}}{\p\zeta}-c_0^2\f{\p}{\p\zeta}\left(\ii k\xi_{r1}+\f{\p\xi_{z1}}{\p\zeta}\right)=F_z.\label{xiz1}
\end{align}
This is a pair of linear equations for the first-order corrections to the amplitude functions, $\xi_{r1}$ and $\xi_{z1}$. As expected, they involve the same linear operator as in the homogeneous equations (\ref{xir0})--(\ref{xiz0}) that occur at leading order, but they are forced by the terms
\begin{align}
  &F_r=2\ii\omega_0\f{\p\xi_{r0}}{\p\tau}-\xi_{z0}\f{\p}{\p\zeta}(2r\Omega_0\Omega_2)+\ii kc_0^2\f{1}{r\rho_0}\f{\p}{\p r}(r\rho_0\xi_{r0})\nonumber\\
  &\qquad-c_0^2\f{\p}{\p r}\left(\f{\rho_0'}{\rho_0}\right),\\
  &F_z=2\ii\omega_0\f{\p\xi_{z0}}{\p\tau}+c_0^2\f{\p}{\p\zeta}\left[\f{1}{r\rho_0}\f{\p}{\p r}(r\rho_0\xi_{r0})\right].
\end{align}
The second term in $F_r$ is due to vertical shear and can be traced to the third term in equation~(\ref{d2xir}). Note that
\begin{equation}
  \f{\p}{\p\zeta}(2r\Omega_0\Omega_2)=\f{\zeta}{H_0^2}\f{\dd c_0^2}{\dd r}.
\end{equation}
An alternative expression for $F_r$ is
\begin{align}
  &F_r=2\ii\omega_0\f{\p\xi_{r0}}{\p\tau}+\ii kc_0^2\f{1}{r\rho_0}\f{\p}{\p r}(r\rho_0\xi_{r0})-\f{\p}{\p r}\left(\f{p_0'}{\rho_0}\right)\nonumber\\
  &\qquad-\left(\ii k\xi_{r0}+\f{\p\xi_{z0}}{\p\zeta}\right)\f{\dd c_0^2}{\dd r}.
\end{align}

The linear operator involved in equations (\ref{xir1})--(\ref{xiz1}) is self-adjoint, having Hermite polynomial eigenfunctions. If we multiply equation~(\ref{xir1}) by $\hat\xi_r^*$ and equation~(\ref{xiz1}) by $\hat\xi_z^*$, add, multiply by $\rho_0$ and integrate vertically through the disc, then the linear operator on the left-hand side can be transferred onto $\hat\xi_r$ and $\hat\xi_z$ by integration by parts, producing zero by virtue of equations (\ref{xir0})--(\ref{xiz0}). In this way we discover the solvability condition for equations (\ref{xir1})--(\ref{xiz1}), which is that the forcing vector is required to be orthogonal to the eigenmode:
\begin{equation}
  \int\rho_0\left(\hat\xi_r^*F_r+\hat\xi_z^*F_z\right)\,\dd\zeta=0.
\label{solvability}
\end{equation}
This condition will determine the slow evolution of the wave amplitude.

\subsection{Amplitude equation}

We omit the uninteresting details of the algebra and vertical integrals involved in manipulating the solvability condition~(\ref{solvability}) into an acceptable form. After the asymptotic scalings are removed, we obtain an evolution equation for the wave amplitude in the form
\begin{equation}
  \f{1}{v_\text{g}}\f{\p\mathcal{A}}{\p t}+\f{\p\mathcal{A}}{\p r}+\f{1}{2}\left[\f{\dd}{\dd r}\ln(\Sigma H^2rv_\text{g})-\left(1-\f{n\Omega^2}{\omega^2}\right)\f{\dd\ln T}{\dd r}\right]\mathcal{A}=0.
\end{equation}

\section{Time-dependent solutions of the wave-amplitude equation}
\label{s:evolving}

If the spatial growth of $\mathcal{A}$ found for a steady wavetrain is factored out of a time-dependent solution of the wave-amplitude equation, what remains is a linear advection equation describing propagation at the group velocity $v_\text{g}(r)$. The solution of that linear advection equation is an arbitrary function of a Lagrangian variable that moves at the local group velocity. Therefore a general, evolving wavetrain can be obtained by multiplying the steady solution (Section~\ref{s:flux_steady}) by an arbitrary function of $(t-t_\text{g}(r))$, where
\begin{equation}
  t_\text{g}(r)=\int_{r_0}^r\f{\dd r'}{v_\text{g}(r')}
\end{equation}
is the group travel time from an arbitrary reference radius $r_0$ to radius~$r$. This procedure works both for the wave amplitude $\mathcal{A}$ and for the wave flux $\mathcal{F}$.

For example, in the case that $n=1$ and $H/r=\epsilon=\cst$ so that $q=-1$, and in units such that $\omega=R=1$ without loss of generality, we have
\begin{equation}
  \mathcal{F}=\f{F(t-t_\text{g})}{r}\exp\left(-\f{1}{3r^3}\right),
\end{equation}
where $F$ is an arbitrary function and
\begin{equation}
  t_\text{g}=-\f{2}{3\epsilon}\left(r^{-3/2}-r^{3/2}\right).
\end{equation}  
Note that $t_\text{g}$ is defined here with reference radius $r_0=1$ and is negative (i.e.\ $-t_\text{g}$ is the time that the wave will take to reach the Lindblad resonance). The steady solution corresponds to $F=\cst$. Suppose instead we apply an initial condition $\mathcal{F}=\mathcal{F}_0(r)$ at $t=0$; then
\begin{equation}
  F(-t_\text{g})=\mathcal{F}_0(r)r\exp\left(\f{1}{3r^3}\right)
\end{equation}
and the relevant solution is
\begin{equation}
  \mathcal{F}=\mathcal{F}_0(r')\f{r'}{r}\exp\left(\f{1}{3r'^3}-\f{1}{3r^3}\right),
\end{equation}
where
\begin{equation}
  r'=\left\{\sqrt{1+\left[\f{3\epsilon(t-t_\text{g})}{4}\right]^2}-\f{3\epsilon(t-t_\text{g})}{4}\right\}^{2/3}.
\end{equation}
The time-dependent relation between $r'$ and $r$ represents the mapping described by outward group propagation: $t_\text{g}(r)=t_\text{g}(r')+t$.

The solution described above is valid for a disc that extends down to $r=0$. If instead the disc has a non-zero inner radius $r_\text{in}$ at which the boundary condition determines the wave amplitude (or flux), then the solution should be modified so that $r'$ is replaced with $r_\text{in}$ whenever $r'<r_\text{in}$. As $t$ increases, the range of $r$ for which $r'<r_\text{in}$ expands. In this way the initial condition is erased and a steady wavetrain, e.g.
\begin{equation}
  \mathcal{F}=\mathcal{F}_\text{in}\f{r_\text{in}}{r}\exp\left(\f{1}{3r_\text{in}^3}-\f{1}{3r^3}\right),
\end{equation}
 is established from the inside out. Here $\mathcal{F}_\text{in}=\mathcal{F}_0(r_\text{in})$ is the specified flux at the inner boundary, which sets the overall amplitude of the steady wavetrain.

\section{Projection onto Hermite polynomials}
\label{s:hermite}

We return to the linearized equations (\ref{Eulin1})--(\ref{Eulin4}) of our model. Appropriately for a thin disc, we adopt the Gaussian vertical profile (\ref{rho_thin}) of the density and assume that $\Omega$ can be approximated as $\Omega_\text{K}(r)$ everywhere except in the last term of equation~(\ref{Eulin2}), for which (cf.\ equation~\ref{omega_thin})
\begin{equation}
  \f{1}{r}\f{\p(r^2\Omega)}{\p z}\approx\f{1}{2\Omega_\text{K}}\f{z}{H^2}\f{\dd c_\text{s}^2}{\dd r}.
\end{equation}
Let us represent the vertical structure of each perturbation variable in the basis of Hermite polynomials of $\zeta=z/H$ (which are orthogonal with respect to an inner product weighted by the Gaussian density distribution):
\begin{align}
  &\f{u_r'}{c_\text{s}}=\sum_{n=0}^\infty a_n(r,t)\,\mathrm{He}_n(\zeta),\\
  &\f{u_\phi'}{c_\text{s}}=\sum_{n=0}^\infty b_n(r,t)\,\mathrm{He}_n(\zeta),\\
  &\f{u_z'}{c_\text{s}}=\sum_{n=1}^\infty c_n(r,t)\,\mathrm{He}_{n-1}(\zeta),\\
  &
  \f{\rho'}{\rho}=\sum_{n=0}^\infty d_n(r,t)\,\mathrm{He}_n(\zeta).
\end{align}
Here $a_n$ to $d_n$ are dimensionless coefficients, dependent on radius and time, to be determined. Then the expression of the linearized equations in the same basis is
\begin{align}
  &\f{1}{\Omega_\text{K}}\f{\p a_n}{\p t}-2b_n=-H\f{\p d_n}{\p r}+\f{\dd H}{\dd r}[nd_n+(n+1)(n+2)d_{n+2}],\label{hermite1}\\
  &\f{1}{\Omega_\text{K}}\f{\p b_n}{\p t}+\f{1}{2}a_n+H\f{\dd\ln c_\text{s}}{\dd r}[c_n+(n+1)c_{n+2}]=0,\\
  &\f{1}{\Omega_\text{K}}\f{\p c_n}{\p t}=-nd_n,\\
  &\f{1}{\Omega_\text{K}}\f{\p d_n}{\p t}+H\left[\f{\p a_n}{\p r}+\f{\dd\ln(r\Sigma c_\text{s})}{\dd r}a_n\right]\nonumber\\
  &\qquad+\f{\dd H}{\dd r}(na_n+a_{n-2})-c_n=0.\label{hermite4}
\end{align}
Here we have used the fact that $\Sigma\propto\rho_\text{m}H$ as well as the properties
\begin{align}
  &\f{\p}{\p r}\,\mathrm{He}_n(\zeta)=-\f{\dd\ln H}{\dd r}\zeta\,\mathrm{He}_n'(\zeta),\\
  &\f{\p}{\p r}\exp\left(-\f{\zeta^2}{2}\right)=\f{\dd\ln H}{\dd r}\zeta^2\exp\left(-\f{\zeta^2}{2}\right),
\end{align}
and then
\begin{align}
  \zeta\,\mathrm{He}_n'(\zeta)&=n\zeta\,\mathrm{He}_{n-1}(\zeta)\nonumber\\
  &=n\,\mathrm{He}_n(\zeta)+n(n-1)\,\mathrm{He}_{n-2}(\zeta),\\
  \zeta^2\,\mathrm{He}_n(\zeta)&=\mathrm{He}_{n+2}(\zeta)+(2n+1)\,\mathrm{He}_n(\zeta)\nonumber\\
  &\qquad+n(n-1)\,\mathrm{He}_{n-2}(\zeta).
\end{align}

Equations (\ref{hermite1})--(\ref{hermite4}) involve some couplings between neighbouring modes of the same parity: the evolution of mode~$n$ is affected by modes $n\pm2$. However, the couplings are weak in the sense that the coefficients of the coupling terms involve radial derivatives of $H$ or~$c_\text{s}$.

To the extent that couplings between different $n$ can be neglected, we have the approximate system
\begin{align}
  &\f{1}{\Omega_\text{K}}\f{\p a_n}{\p t}-2b_n=-H\f{\p d_n}{\p r}+\f{\dd H}{\dd r}nd_n,\\
  &\f{1}{\Omega_\text{K}}\f{\p b_n}{\p t}+\f{1}{2}a_n+H\f{\dd\ln c_\text{s}}{\dd r}c_n=0,\\
  &\f{1}{\Omega_\text{K}}\f{\p c_n}{\p t}=-nd_n,\\
  &\f{1}{\Omega_\text{K}}\f{\p d_n}{\p t}+H\left[\f{\p a_n}{\p r}+\f{\dd\ln(r\Sigma c_\text{s})}{\dd r}a_n\right]+\f{\dd H}{\dd r}na_n-c_n=0
\end{align}
for each mode separately.

Consider a short-wavelength limit in which the radial derivative $\p/\p r$ acting on a wave amplitude coefficient such as $a_n$ is replaced by a multiplication by $\ii k$, where $k$ is the radial wavenumber such that $kr\gg1$, and in which the time-dependence is through a common factor of $\exp(-\ii\omega t)$. Then, to a first approximation, we obtain the algebraic system
\begin{align}
  &-\f{\ii\omega}{\Omega}a_n-2b_n\approx-\ii kHd_n,\\
  &-\f{\ii\omega}{\Omega}b_n+\f{1}{2}a_n+\f{1}{2}q\epsilon c_n\approx0,\\
  &-\f{\ii\omega}{\Omega}c_n=-nd_n,\\
  &-\f{\ii\omega}{\Omega}d_n+\ii kHa_n-c_n\approx0,
\end{align}
in which we write $\Omega$ for $\Omega_\text{K}$, $q$ for $\dd\ln T/\dd\ln r$ and $\epsilon$ for $H/r$.
This system leads to the local dispersion relation
\begin{equation}
  \left(\f{\omega^2}{\Omega^2}-1\right)\left(\f{\omega^2}{\Omega^2}-n\right)+\ii nq\epsilon kH=\f{\omega^2}{\Omega^2}(kH)^2.
\label{ldr_hermite_appendix}
\end{equation}

\bsp    
\label{lastpage}
\end{document}